\PassOptionsToPackage{prologue,dvipsnames,svgnames}{xcolor}
\documentclass[11pt]{article}
\usepackage{fullpage}

\usepackage{amsmath,amssymb}
\usepackage{bm}
\usepackage[round]{natbib}

\usepackage{amssymb}
\usepackage{mathrsfs}
\usepackage{graphicx}
\usepackage{subcaption}
\usepackage{pdfpages}
\usepackage{afterpage}
\usepackage{caption}
\usepackage{multirow}
\usepackage{array}
\usepackage{rotating}
\usepackage[flushleft]{threeparttable}
\usepackage{multirow}
\usepackage{enumitem} 
\usepackage{algorithm2e}
\usepackage{svg}
\usepackage{stackengine}
\usepackage{float}
\RestyleAlgo{ruled}
\SetAlgorithmName{Procedure}{procedure}{List of Procedure}
\usepackage{tikz} 
\usetikzlibrary{matrix}
\usepackage{adjustbox}

\usepackage{preamble/star}
\usepackage[group-separator={,},group-minimum-digits={3}]{siunitx}

\usepackage[colorlinks,linkcolor=blue,anchorcolor=blue,citecolor=blue,urlcolor=black]{hyperref}
\usepackage{cleveref}
\svgpath{{Figure/}}

\newcolumntype{P}[1]{>{\centering\arraybackslash}p{#1}}

\def\##1\#{\begin{align}#1\end{align}}
\def\$#1\${\begin{align*}#1\end{align*}}

\newcommand{\Rom}[1]{\text{\uppercase\expandafter{\romannumeral #1\relax}}}

\usepackage{color}

\usepackage{array} 
\newcolumntype{C}[1]{>{\centering\arraybackslash}p{#1}}  

\definecolor{wjs}{RGB}{200,0,50}

\begin{document}



\title{How to Find Fantastic AI Papers: Self-Rankings as a Powerful Predictor of Scientific Impact Beyond Peer Review}

\author{Buxin Su\thanks{University of Pennsylvania; Email: \texttt{\{subuxin,ncollina,bxzhao,suw\}@upenn.edu}.} \and Natalie Collina\footnotemark[1] \and Garrett Wen\thanks{Yale University; Email: \texttt{gang.wen@yale.edu}.} \and Didong Li\thanks{University of North Carolina at Chapel Hill; Email: \texttt{didongli@unc.edu}.} \and Kyunghyun Cho\thanks{New York University; Email: \texttt{kyunghyun.cho@nyu.edu.}} \and Jianqing Fan\thanks{Princeton University; Email: \texttt{jqfan@princeton.edu}.} \and Bingxin Zhao\footnotemark[1]~\thanks{Corresponding authors.} \and Weijie Su\footnotemark[1]~\footnotemark[6]}

\date{\today}

\maketitle

\begin{abstract}

Peer review in academic research aims not only to ensure factual correctness but also to identify work of high scientific potential that can shape future research directions. This task is especially critical in fast-moving fields such as artificial intelligence (AI), yet it has become increasingly difficult given the rapid growth of submissions. In this paper, we investigate an underexplored measure for identifying high-impact research: authors' own rankings of their multiple submissions to the same AI conference. Grounded in game-theoretic reasoning, we hypothesize that self-rankings are informative because authors possess unique understanding of their work's conceptual depth and long-term promise. To test this hypothesis, we conducted a large-scale experiment at a leading AI conference, where 1,342 researchers self-ranked their 2,592 submissions by perceived quality. Tracking outcomes over more than a year, we found that papers ranked highest by their authors received twice as many citations as their lowest-ranked counterparts; self-rankings were especially effective at identifying highly cited papers (those with over 150 citations). Moreover, we showed that self-rankings outperformed peer review scores in predicting future citation counts. Our results remained robust after accounting for confounders such as preprint posting time and self-citations. Together, these findings demonstrate that authors' self-rankings provide a reliable and valuable complement to peer review for identifying and elevating high-impact research in AI.

\end{abstract}


\section{Introduction}
\label{sec:1}



A central objective of peer review is to identify and promote research of high impact that will shape the future of a field. Yet, this process is under unprecedented strain, particularly in rapidly expanding areas such as artificial intelligence (AI), which face an explosion in paper submissions \citep{lipton2019troubling, sculley2018avoiding, papercopilot2025statsicml}. For the two largest AI conferences, submissions to the International Conference on Machine Learning (ICML) rose from \num{1676} in 2017 to \num{12107} in 2025, while those to the Conference on Neural Information Processing Systems (NeurIPS) increased from \num{3240} in 2017 to \num{21575} in 2025. This exponential growth has placed a heavy burden on peer review at AI conferences, leading to reliance on graduate and undergraduate students as reviewers \citep{stelmakh2021novice}, many without prior publications at these venues, and increasingly, the use of language models for review \citep{liang2024can}. This raises serious concerns about the quality of the review. A striking example is the NeurIPS 2021 experiment, which revealed that approximately half of the accepted papers would have been rejected under a second independent review \citep{cortes2021inconsistency,beygelzimer2023has}. This inconsistency highlights the challenge of identifying research with lasting impact amidst the overwhelming volume of incremental work. A consequence of the decline of peer review that we cannot preclude is that the advancement of AI may follow a less-than-optimal path.

While the supply of qualified reviewers is strained, far less research has focused on eliciting expert assessment from the authors themselves \citep{aziz2019strategyproof, mattei2020peernomination, srinivasan2021auctions,su2021you,rastogi2022authors}. Authors possess an unparalleled understanding of their own work, encompassing its theoretical foundations, limitations, and the subtle details fundamental to its long-term scientific impact. This deep insight stands in contrast to the focus of many overburdened reviewers who, under tight deadlines, may prioritize quantifiable gains over a submission's broader scientific implications. This is particularly true in AI research, where many reviewers are junior researchers who often focus on marginal improvements in predictive accuracy (such as a benchmark test accuracy increasing from 91\% to 92\%) and may lack the experience to evaluate a paper's long-term significance \citep{shah2018design}.

To empirically test the hypothesis that author insight predicts scientific impact, a primary obstacle is soliciting candid self-assessments, as authors may be inclined to inflate the quality of their work. One approach is to elicit a self-ranking of submissions from the same author, which is practical as it is common for the same author to submit multiple papers to the same conference in AI research. This design elicits comparative rather than absolute judgments, so authors cannot simply claim all their submissions are of the highest quality. Under certain conditions, this comparative design incentivizes authors to truthfully report their true rankings \citep{su2021you,su2025you,yan2023isotonic}. To empirically quantify the predictive power of authors' self-rankings, we designed and implemented a large-scale experiment at ICML in 2023, with the approval of the conference organizers. We asked authors who had submitted multiple papers to rank them according to their perceived quality and significance. The experiment yielded a rich dataset of self-rankings from 1,342 AI researchers, encompassing 2,592 submissions to ICML 2023, along with their official review scores and acceptance decisions.

Our analysis of the ICML 2023 data reveals that authors' self-rankings are a powerful predictor of a paper's future impact, measured by citations accumulated over 16 months. Citation count is a widely used, albeit imperfect, metric for the scientific impact of research \citep{aksnes2019citations,huang2022evaluating}. Submissions ranked highest by their authors received, on average, twice as many citations as those they ranked lowest, a trend that held for both accepted and rejected submissions, suggesting that self-rankings can not only denoise review scores but also capture a distinct dimension of impact \citep{su2025icml}. The predictive power was particularly striking for identifying high-impact work: of the 22 papers in our dataset that garnered over 150 citations, 17 were ranked highest by their authors. For comparison, we demonstrate that these self-rankings are a more accurate predictor of future citations than the review scores. These findings are highly statistically significant and remained robust after controlling for potential confounding factors such as early dissemination on preprint servers and authors' self-citations. Our results suggest that authors' self-rankings are a valuable, cost-effective signal that could be integrated into future review systems to help identify and elevate research of the highest impact.

\begin{figure}[!htp]
    \centering
    \includegraphics[width=0.95\textwidth]{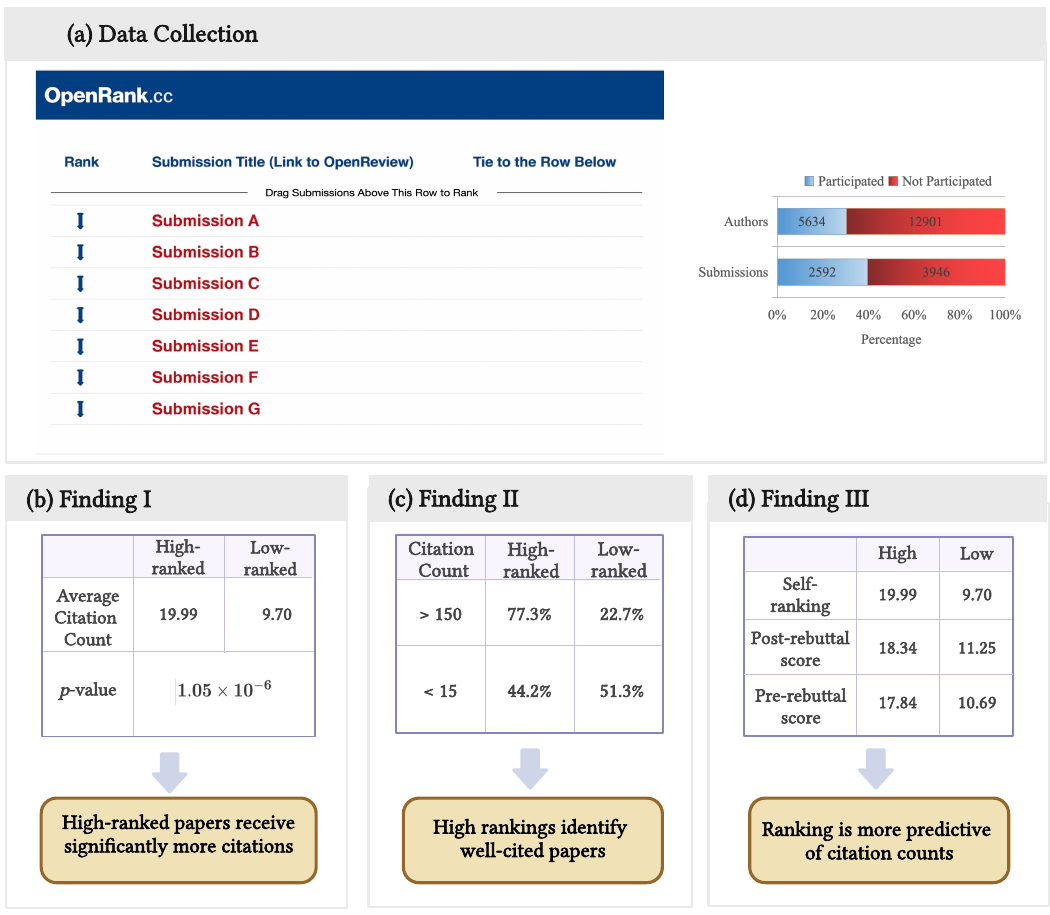}
    \caption{(a) The illustration for the Phase One survey experiment, with summary statistics (see Section~\ref{sec:summery} for details and Figure~\ref{fig:survey_interface} for the real interface). (b) Comparison of citation counts between high- and low-ranked submissions, revealing a statistically significant difference ($P = 1.05 \times 10^{-6}$; see Section~\ref{par:rank_vs_avg_ciation}). (c) Proportions of high- and low-ranked submissions, categorized by how well they are cited (i.e., well-cited vs.\ less-cited) (see Section~\ref{par:rank_idfy_good}). (d) Mean citation counts for submissions partitioned into high- and low-value groups according to three metrics: self-rankings, pre-rebuttal scores, and post-rebuttal scores. A larger difference in means indicates greater predictive power for the metric (see Section~\ref{sec:rank_vs_score}).
    }
    \label{fig:overview}
\end{figure}

\section{Data collection}
\label{sec:summery}
Our experiment was conducted in two phases. In Phase One, we carried out a survey-based experiment on OpenReview, which hosted the ICML 2023 peer review process for \num{6538} submissions from \num{18535} authors. The experiment itself was implemented with \href{https://openrank.cc/}{OpenRank.cc}, a platform we developed for this purpose. On January 26, 2023, immediately after the submission deadline, an official email was distributed via OpenReview to all ICML authors requesting information about their submissions. Authors with multiple submissions were asked to rank their papers based on perceived quality, along with answering several questionnaire items. 
In total, \num{5634} authors completed the survey (response rate: $30.4\%$), of which \num{1342} authors with multiple submissions provided rankings. Altogether, \num{2592} submissions were ranked by at least one author, accounting for $39.6\%$ of all submissions. 
It is common for AI researchers to submit multiple papers to the same conference. Indeed, at ICML 2023, \num{4505} of the \num{18535} authors had more than one submission, and \num{5035} of the total \num{6538} submissions had at least one author having more than one submission. 
A paper submitted to an AI conference is evaluated by multiple reviewers. Before the rebuttal phase, every reviewer assigns a pre-rebuttal review score—at ICML 2023, on a 1–10 scale. During the rebuttal period, authors may respond to reviewers’ comments and clarify points without making substantial changes to the original submission. After considering the rebuttal, reviewers may revise their scores; these revised scores are referred to as post-rebuttal scores. We obtained the pre- and post-rebuttal scores, as well as the final decisions for all submissions, from OpenReview.

In Phase Two of the experiment, we used Semantic Scholar to collect citation data for each ranked submission.\footnote{Semantic Scholar provides citation records for a paper within a specified time period. See Section~\ref{sec:other_metric} for results based on Google Scholar.} For each submission, we retrieved all papers that cited the submission between July 23, 2023 and November 22, 2024.
For each citing paper, we extracted the title, author list, and publication date. A key challenge was accounting for multiple submission versions, which often involved revised titles or author lists before or after ICML 2023. Additionally, some titles or author names include special characters, such as ``Schr{\"o}dinger,'' which introduced minor mismatches during automated title matching. To ensure accuracy, we retained only submissions with an exact match in both title and author list on Semantic Scholar (see Section \ref{sec:method} for details). After filtering, we obtained a final dataset consisting of 797 authors with multiple ranked submissions, each with valid citation data, involving a total of 1,527 unique submissions. All subsequent citation analyses are based on this filtered set. Across these submissions, the average number of citations per paper during the study window was \num{14.73}. Half of these submissions received at least \num{4} citations, and one-quarter received at least \num{11} citations. The most-cited submission received \num{979} citations.\footnote{These statistics are based on the \num{1527} submissions with valid citation counts from Semantic Scholar.}


\begin{figure}[!htp]
    \centering
    \begin{subfigure}[b]{0.495\textwidth}
        \includegraphics[width=\textwidth]{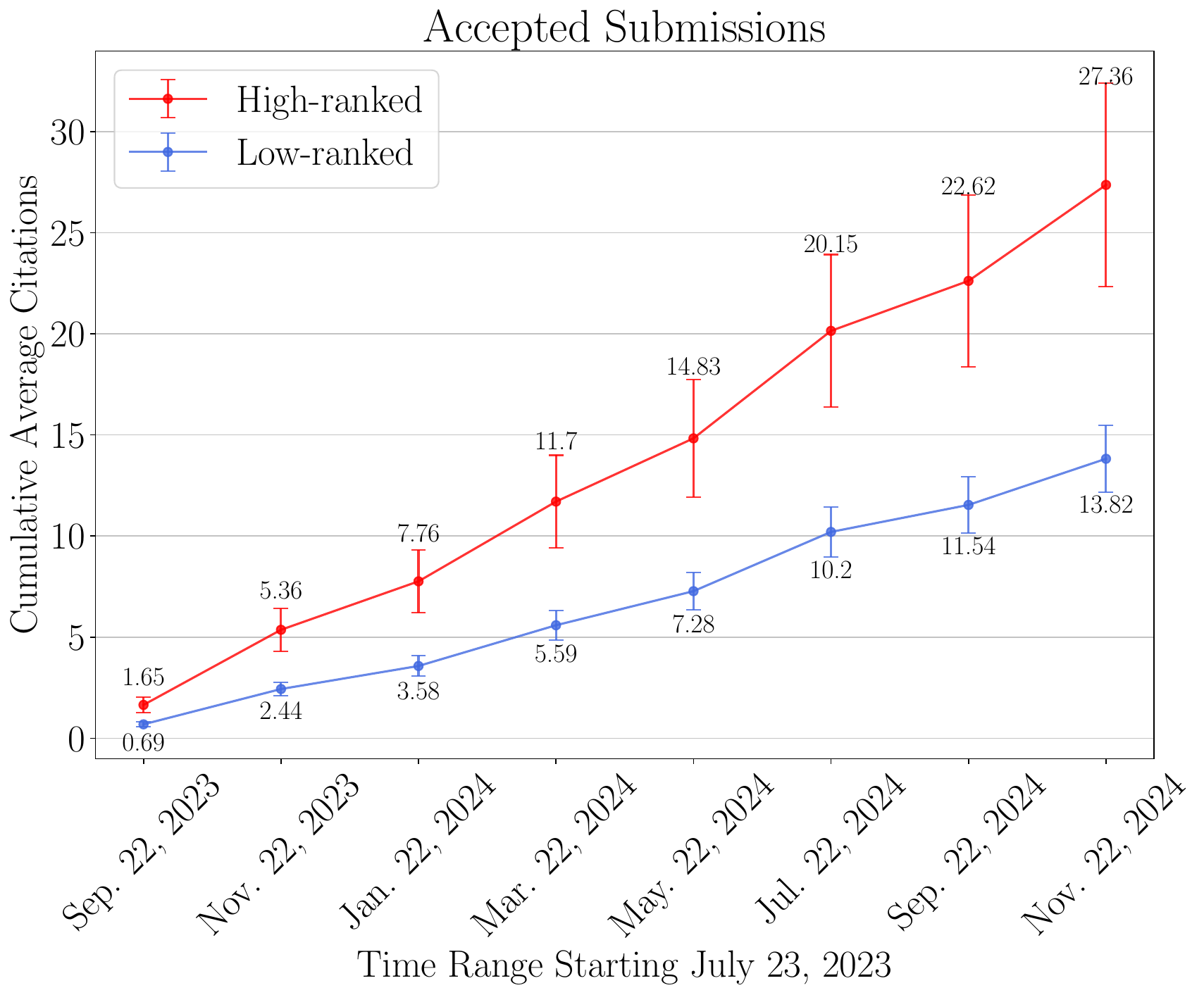}
    \end{subfigure}
    \hfill
    \begin{subfigure}[b]{0.495\textwidth}
        \includegraphics[width=\textwidth]{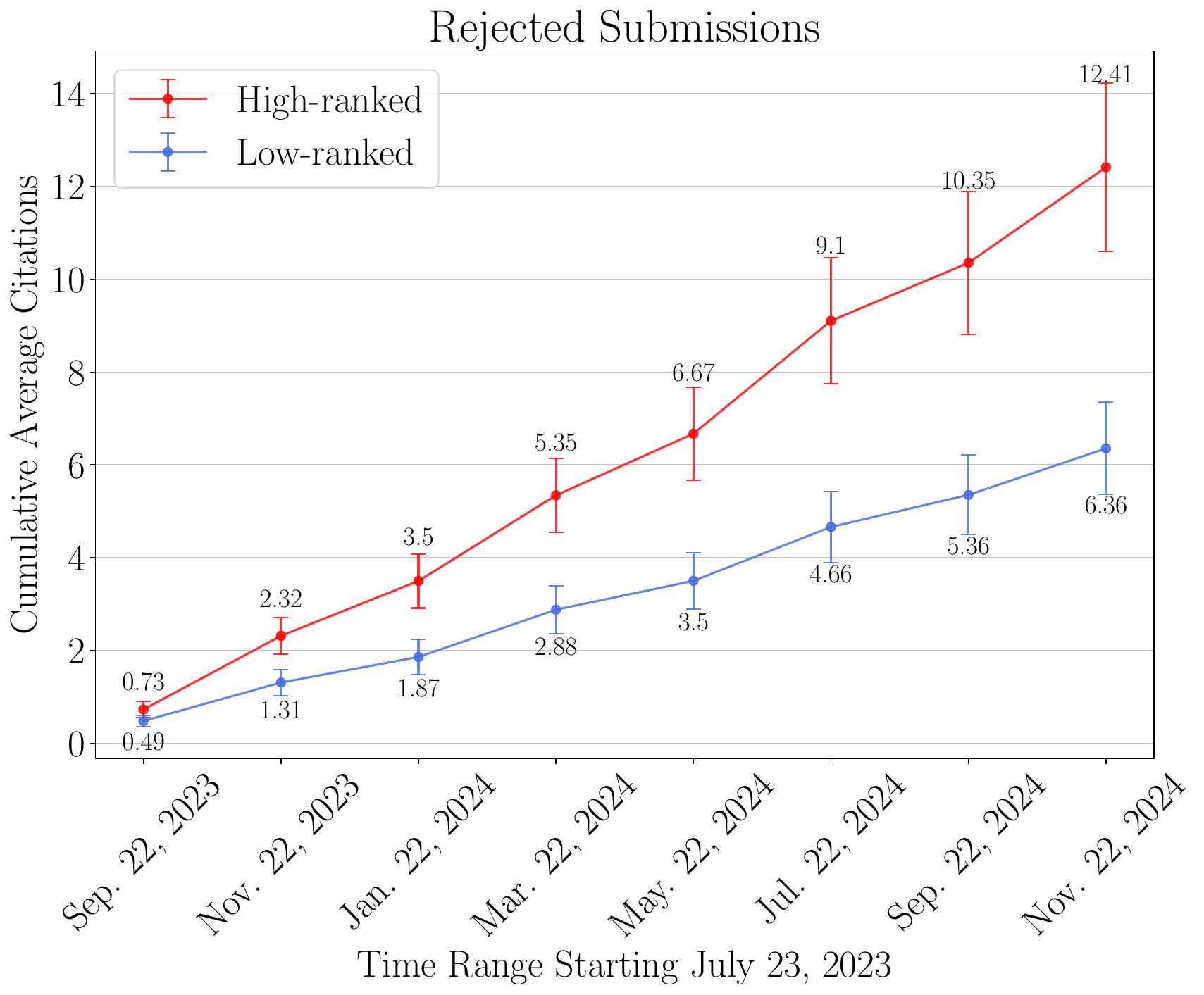}
    \end{subfigure}
    \hfill
    \begin{subfigure}[b]{0.49\textwidth}
        \includegraphics[width=\textwidth]{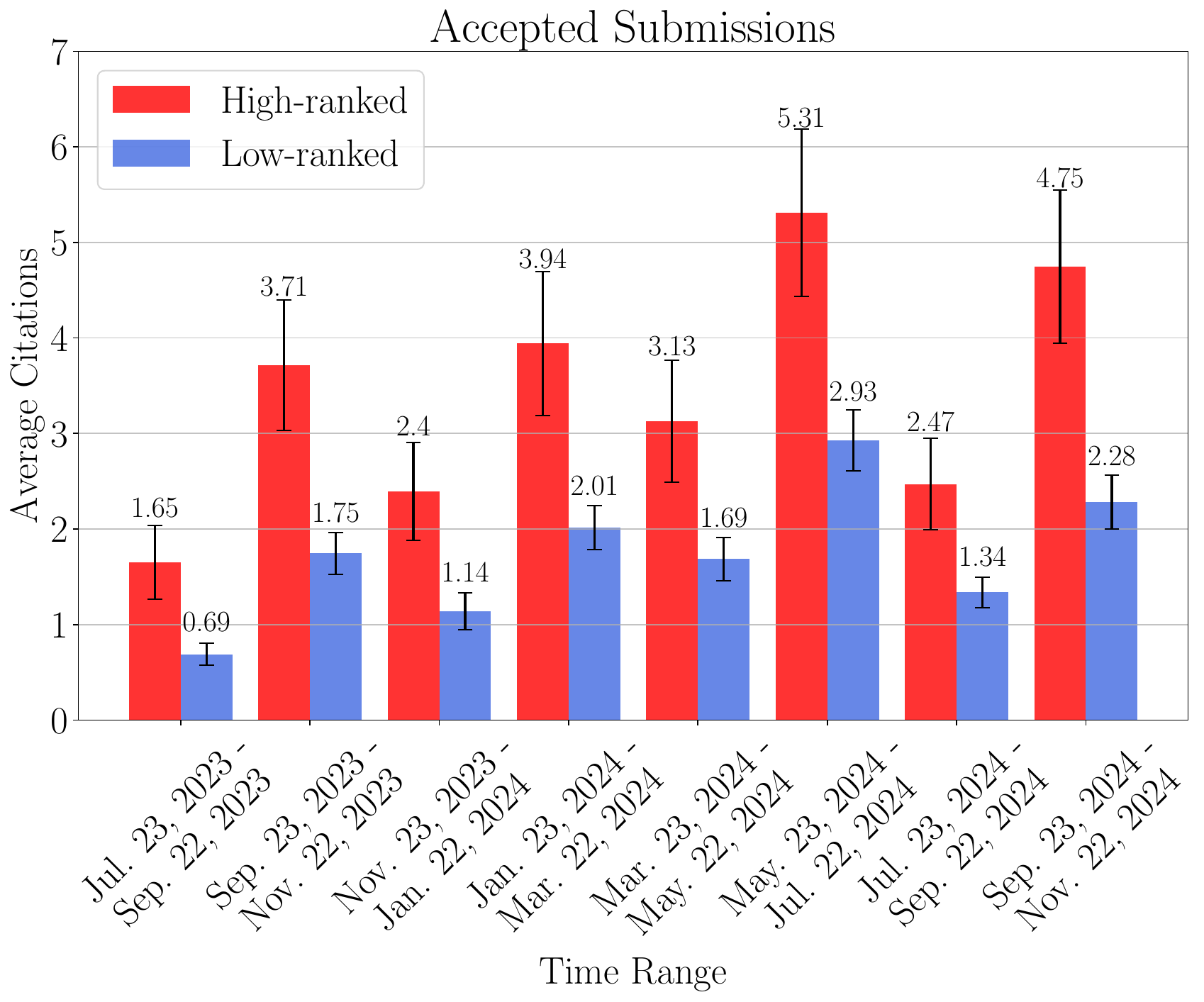}
    \end{subfigure}
    \hfill
    \begin{subfigure}[b]{0.49\textwidth}
        \includegraphics[width=\textwidth]{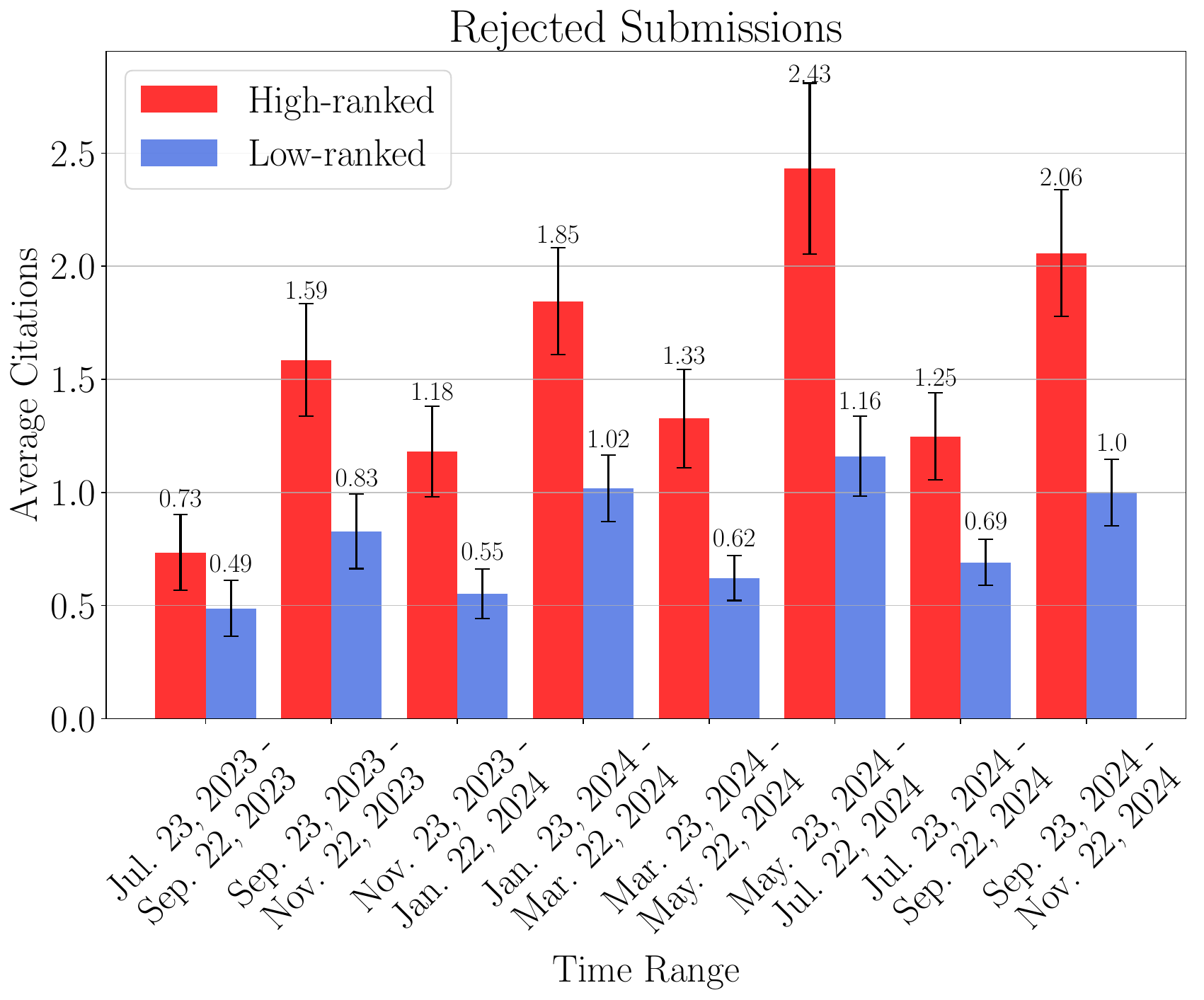}
    \end{subfigure}
    \caption{{High-ranked papers received about twice as many citations as low-ranked ones.} (Cumulative) average number of citations in each two-month period from July 23, 2023 to November 22, 2024, comparing high-ranked papers (red bars) and low-ranked papers (blue bars), conditional on final accept/reject decisions. {Left panel}: accepted submissions. {Right panel}: rejected submissions. Across all intervals and decision categories, high-ranked papers consistently received substantially more citations than low-ranked papers.}
    \label{fig:norm_citation}
\end{figure}

\section{High-ranked papers received twice as many citations as low-ranked papers}
\label{par:rank_vs_avg_ciation}

We began by investigating the relationship between authors' self-rankings and citation counts. For each author, we categorized the top-ranked submission (ranked 1) as high-ranked and the lowest-ranked submission as low-ranked. As shown in Figure~\ref{fig:overview}(b), high-ranked papers received an average of 19.99 citations, which is double the average for low-ranked papers ($P=1.05 \times 10^{-6}$, two-sided $t$-test). We next conducted a more detailed analysis by conditioning on final decisions and time period. Among the 797 authors with multiple submissions, 280 authors had multiple accepted submissions and 343 had multiple rejected submissions.\footnote{We regard ``Accepted as Poster'' or ``Accepted as Poster \& Oral'' as accepted, and ``Rejected'' or ``Withdrawn or Desk Rejected'' as rejected.} Within each author's accepted submissions, we partitioned them into the ``accepted and high-ranked'' group and the ``accepted and low-ranked'' group. A similar classification was applied to rejected submissions, resulting in four groups. Figure~\ref{fig:norm_citation} reports average citation counts for each group over successive two-month intervals from July 23, 2023 to November 22, 2024.

The upper panel of Figure~\ref{fig:norm_citation} shows that cumulative citation counts grew approximately linearly over time, with high-ranked papers receiving citations at a rate about twice that of low-ranked papers. Between July 23, 2023, and November 22, 2024, accepted and rejected high-ranked papers received, on average, 27.36 and 12.41 citations, respectively, which were 1.98 and 1.95 times more than the corresponding low-ranked groups. The lower panel of Figure~\ref{fig:norm_citation} provides further evidence that, in every two-month period, high-ranked papers consistently obtained more citations than low-ranked papers. Moreover, when comparing citation counts over time, we identified four two-month periods with noticeably higher volumes: September--November 2023, January–March 2024, September--November 2024, and May--July 2024, in increasing order. These time intervals correspond to the submission deadlines of several major AI conferences: International Conference on Learning Representations (ICLR) 2023, ICML 2023, ICLR 2024, and NeurIPS 2023, respectively. Notably, the order of citation volume across these intervals aligns with the submission volume at these conferences: 4,874 (ICLR 2023), 6,538 (ICML 2023), 7,262 (ICLR 2024), and 12,345 (NeurIPS 2023).

As shown in Figure~\ref{fig:top_k_citation} in the Supplementary Material, authors' self-rankings beyond the first and last positions were also informative of citation counts. Specifically, we analyzed how average citation counts vary across rankings for two groups of authors: 76 authors with more than three accepted submissions and 86 authors with more than three rejected submissions, all of whom provided valid rankings. We found that, across both decision categories, low-ranked submissions tended to receive fewer citations.



\section{Self-rankings identify well-cited papers on the edge}
\label{par:rank_idfy_good}

While we have shown that higher-ranked papers received substantially more citations on average, it is also important to assess predictivity in the edge case of highly cited papers.
Building on the earlier observation, we next examined how well-cited submissions are distributed across authors' self-rankings. 
As shown in Figure~\ref{fig:overview}(c), among all submissions that received more than \num{150} citations, $77.27\%$ were ranked first by at least one of their authors. Specifically, of the 22 submissions exceeding this threshold, 17 were ranked first, 3 were ranked second, and 5 were ranked last by at least one author. Only one submission was ranked both first and last by different authors, while another was neither ranked first nor last.
We chose the top 22 submissions because $22/1527 \approx 1.44\%$, which roughly corresponds to the proportion of ``Accepted as Oral'' at major AI conference. 
Moreover, most well-cited papers belong to the high-ranked category. Focusing on high-ranked papers alone suffices to identify most well-cited papers.

\begin{figure}[!htp]
    \centering
    \begin{subfigure}[b]{0.43\textwidth}
        \includegraphics[width=\textwidth]{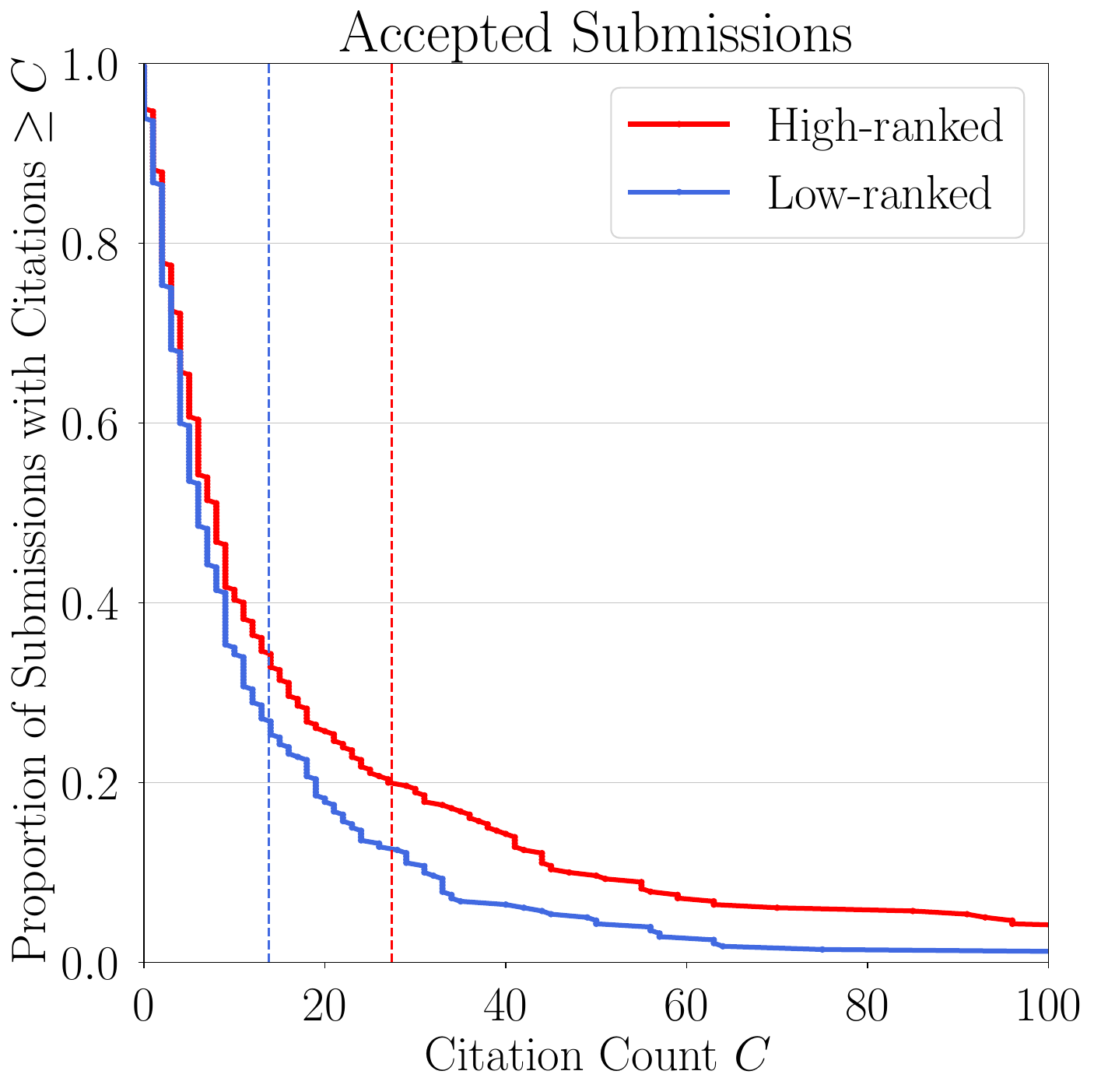}
    \end{subfigure}
    \hspace{0.5cm}
    \begin{subfigure}[b]{0.43\textwidth}
        \includegraphics[width=\textwidth]{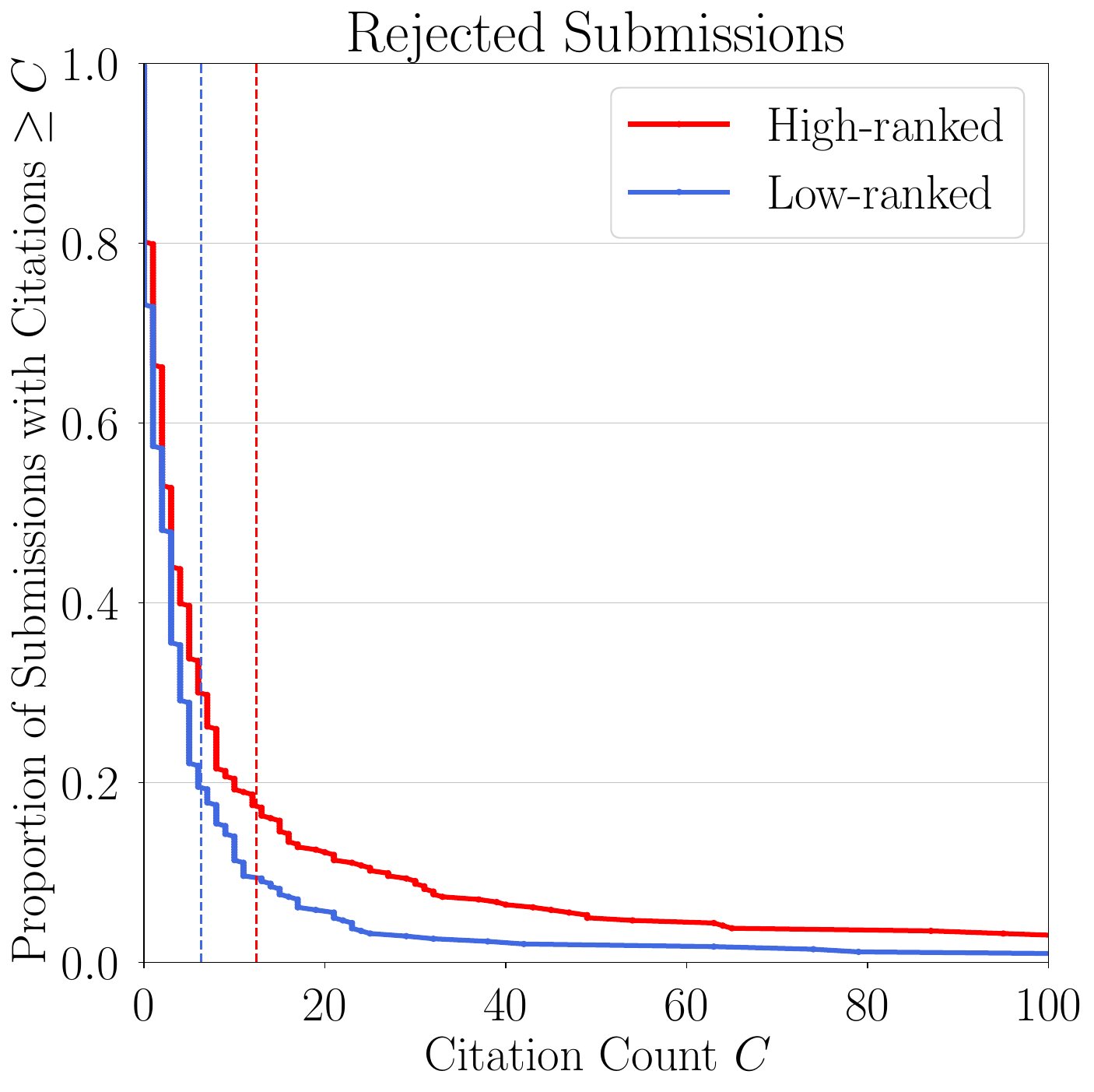}
    \end{subfigure}
    \caption{
    {Most of the well-cited papers fall into the high-ranked category.} Empirical complementary cumulative distribution of citation counts, showing the proportion of papers with more than $C$ citations. The left panel presents accepted submissions; the right panel shows rejected ones. At every citation threshold $C$, a higher proportion of high-ranked papers exceed $C$ than low-ranked papers. Average citation counts for high- and low-ranked submissions are highlighted using vertical dashed lines.}
    \label{fig:CCDF_versus_citation}
\end{figure}

Figure~\ref{fig:CCDF_versus_citation} displays the empirical complementary cumulative distribution functions of citation counts, with vertical dashed lines indicating the average citation counts of high- and low-ranked submissions. For any citation threshold $C$, a greater fraction of high-ranked submissions exceeded $C$ compared to their low-ranked counterparts. The largest gap occurred at $C = 35$, where 16.79\% of high-ranked papers exceeded this threshold, in contrast to 6.79\% of low-ranked papers. Furthermore, the tail of the high-ranked group is noticeably heavier, indicating a substantially larger proportion of well-cited papers. For instance, 4.18\% of high-ranked submissions received more than 100 citations, approximately three times the corresponding proportion among low-ranked papers. Table~\ref{tab:citation_versus_rank} in the Supplementary Material summarizes the percentile of citation counts and further supports this pattern. Among accepted submissions, the most-cited high-ranked paper accumulated 979 citations, nearly three times the maximum of any low-ranked papers. We also observed that high-ranked papers consistently received more citations than low-ranked papers, which supports the observations in Figure \ref{fig:norm_citation}.

\section{Self-rankings outperform review scores in predicting citations}
\label{sec:rank_vs_score}


We further compare the predictive power of authors’ self-rankings versus reviewer scores for submissions’ future citation counts. Conditioning on the final decision, we analyze 280 authors with multiple accepted and ranked submissions and 343 authors with multiple rejected and ranked submissions. As shown in Table~\ref{tab:rank_vs_score} and Figure~\ref{fig:overview}(d), the citation-count gap between high and low category is largest when submissions are grouped by self-rankings, compared with reviewer scores, both pre- and post-rebuttal. In particular, only authors’ self-rankings yield statistically significant differences in citation counts between high- and low-ranked papers (paired $t$-test, $P = 9.34\times10^{-3}$). This indicates that self-rankings are the most predictive of submissions’ citation counts. These patterns are even more particularly pronounced among rejected submissions, possibly because reviewers may not differentiate scores as carefully once a paper falls below the acceptance threshold. Notably, the most-cited paper in our dataset, which accumulated 979 citations between July 23, 2023 and November 22, 2024, was ranked first by its authors but received a post-rebuttal score of 5, lower than that of their other submissions, and was therefore accepted only as a poster.


\begin{table}[!htp]
\centering
\renewcommand{\arraystretch}{1.2}
\resizebox{\textwidth}{!}{
\begin{tabular}{>{\centering\arraybackslash}m{1.7cm}|p{1.3cm}||c|c|c|c|c|c}
\hline \hline
 & & \multicolumn{3}{c}{Citation Count for Accepted Papers} & \multicolumn{3}{|c}{Citation Count for Rejected Papers} \\
\cline{3-5} \cline{6-8}
 & & \begin{tabular}{@{}c@{}} Self-Ranking \end{tabular} & \begin{tabular}{@{}c@{}} Post-Rebuttal \\ Score \end{tabular} & 
\begin{tabular}{@{}c@{}} Pre-Rebuttal \\ Score \end{tabular} & \begin{tabular}{@{}c@{}} Self-Ranking \end{tabular} & \begin{tabular}{@{}c@{}} Post-Rebuttal \\ Score \end{tabular} & 
\begin{tabular}{@{}c@{}} Pre-Rebuttal \\ Score \end{tabular} \\
\hline
\multirow{2}{*}{\centering Mean} & High & 27.36 & 25.58 & 23.62 & 12.41 & 9.99 & 9.98 \\
\cline{2-8}
 & Low  & 13.82 & 17.64 & 18.18 & 6.36  & 9.15 & 8.70 \\
\hline
\multicolumn{2}{c||}{$p$-value}  & $9.34 \times 10^{-3}$ & $0.13$ & $0.30$ & $7.7 \times 10^{-4}$ & $0.65$ & $0.49$ \\
\hline
\multirow{2}{*}{\centering Max} & High & 979.0 & 768.0 & 768.0 & 253.0 & 253.0 & 253.0 \\
\cline{2-8}
 & Low  & 331.0 & 979.0 & 979.0 & 205.0 & 251.0 & 205.0 \\
\hline
\multirow{2}{*}{\centering \begin{tabular}{@{}c@{}} 75th \\ percentile \end{tabular}} & High & 21.0 & 21.0 & 18.0 & 8.0 & 7.0 & 7.0 \\
\cline{2-8}
 & Low  & 15.0 & 15.0 & 16.25 & 5.0 & 6.0 & 6.0 \\
\hline
\end{tabular}}
\caption{Authors' self-rankings better identify impactful papers. Average number of citations (from July 23, 2023 to November 22, 2024) for (1) high-ranked versus low-ranked papers, (2) high-scored versus low-scored papers in the post-rebuttal phase, and (3) high-scored versus low-scored papers in the pre-rebuttal phase, grouped by their final decisions in ICML 2023. At a significance level of $P = 0.05$, only the high-ranked papers show a significantly higher number of citations compared to the low-ranked papers, based on a paired $t$-test. In contrast, the differences in citation counts between high and low groups based on pre-rebuttal and post-rebuttal scores are not statistically significant.
}
\label{tab:rank_vs_score}
\end{table}

\begin{figure}[!htp]
    \centering
    \begin{subfigure}[b]{0.49\textwidth}
        \includegraphics[width=\textwidth]{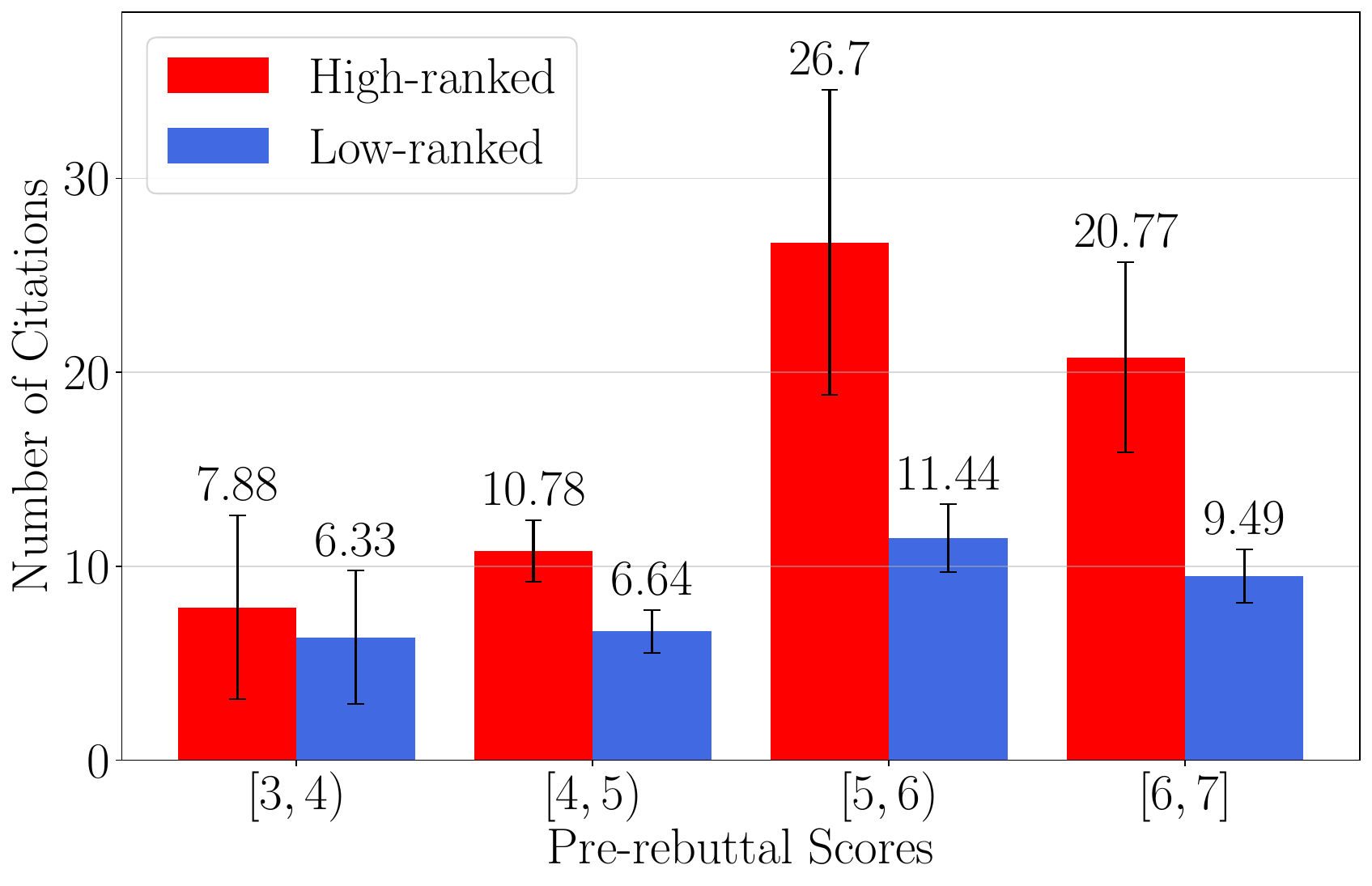}
    \end{subfigure}
    \begin{subfigure}[b]{0.49\textwidth}
        \includegraphics[width=\textwidth]{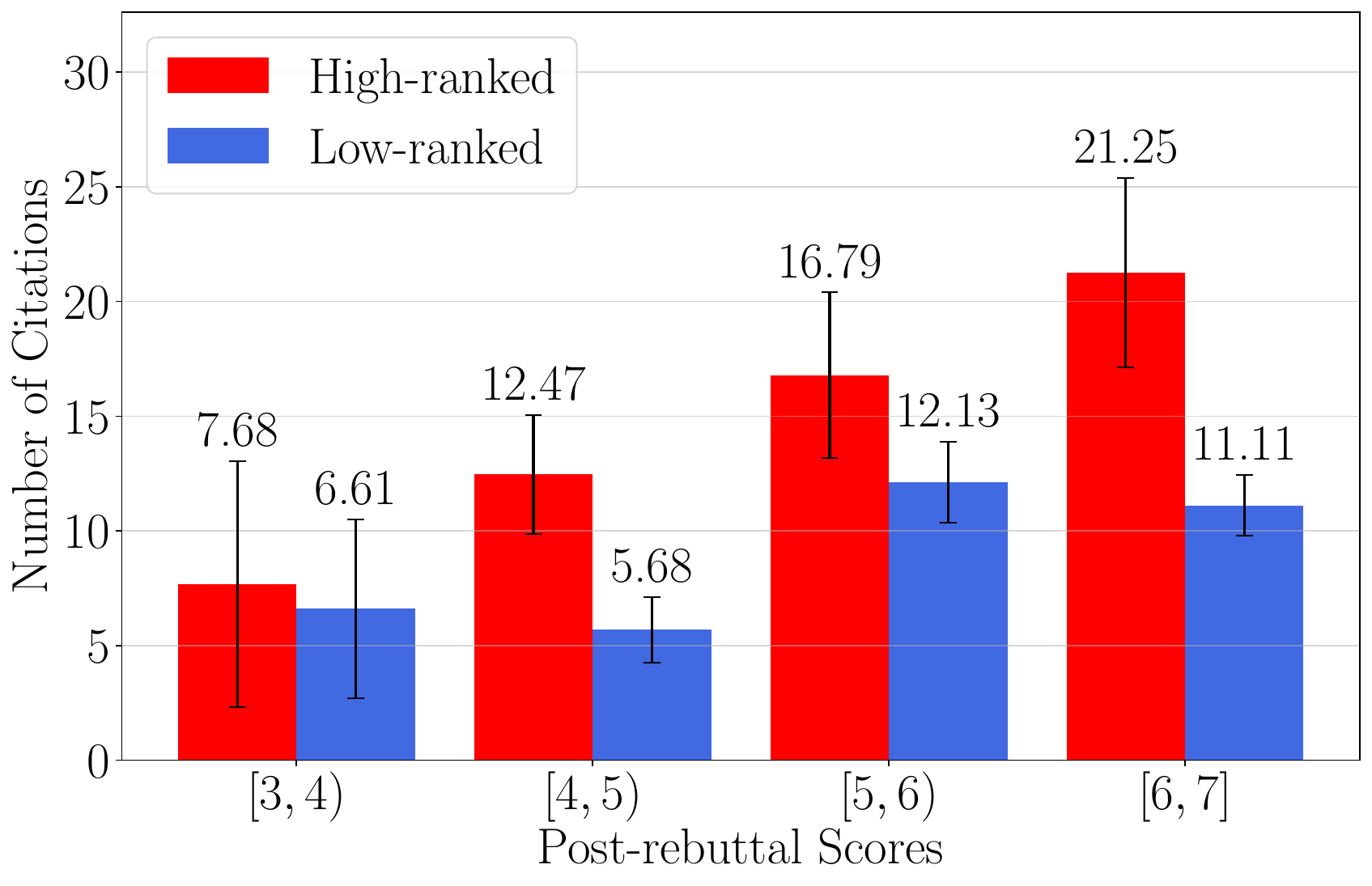}
    \end{subfigure}
    \caption{Average number of citations from July 23, 2023 to November 22, 2024 for high- vs.\ low-ranked papers, grouped by pre-rebuttal and post-rebuttal review scores in ICML 2023. Among submissions with similar review scores, high-ranked papers still received significantly more citations than low-ranked ones. In particular, review scores failed to identify the most impactful papers, whereas authors' self-rankings succeeded in doing so.}
    \label{fig:rank_vs_score}
\end{figure}

Figure~\ref{fig:rank_vs_score} further validates the findings by stratifying submission pairs into four groups with similar pre- or post-rebuttal scores. For example, an author ranks a submission $A$ as better than $B$, and this pair will only be included if their review scores are close, for example, both between 5 and 6. Figure~\ref{fig:summery_cond_score} in the Supplementary Material shows the number of authors with multiple ranked submissions within each review score group.
Even within groups of similar review scores, high-ranked papers by authors tended to accumulate more citations than their low-ranked counterparts. Notably, submissions with pre-rebuttal scores in the 5 to 6 range received the highest citation counts on average, driven by the two most-cited papers in our dataset, both ranked first by their authors. Moreover, the citation gap between high- and low-ranked papers widens as review scores increase. This suggests that while review scores correlates with paper impact, they may not be able to identify the most impactful work. 
In particular, in high-score regimes, authors' self-rankings provide more complementary information that helps capture high-impact papers.

\begin{table}[!htp]
\centering
\begin{tabular}{m{0.17\textwidth}|m{0.17\textwidth}|m{0.17\textwidth}|m{0.17\textwidth}}
\hline\hline
 \centering Self-Ranking & \centering Pre-rebuttal score & \centering Post-rebuttal score & \centering Final Decisions \tabularnewline
\hline
\centering  \centering 17 (high-ranked) & \centering 14 (high-scored) & \centering 14 (high-scored) & \centering 15 (accepted) \tabularnewline
\hline
\centering 5 (low-ranked) & \centering 6 (low-scored) & \centering 8 (low-scored)  & \centering 7 (rejected) \tabularnewline
\hline
\end{tabular}
\caption{Distribution of 22 submissions with more than 150 citations. The largest proportion falls into the high-ranked group, while the smallest proportion falls into the low-ranked group, suggesting that authors' self-rankings are better at identifying well-cited papers than review scores or final decisions. 
Note that among 22 submissions that received more than 150 citations, 14 having highest pre-rabuttal scores among all of an author’s submissions, 6 having lowest pre-rebuttal scores. One submission having both highest and lowest pre-rebuttal scores by different authors and 3 submissions having neither highest or lowest pre-rebuttal scores among all of their authors. 
}
\label{tab:well-cited}
\end{table}

As in Section~\ref{par:rank_vs_avg_ciation}, in addition, we place the highest-scored submissions in one group and the lowest-scored submissions in the other. We first observe that authors' self-rankings are more effective than either review scores or final decisions in identifying well-cited papers. In Table~\ref{tab:well-cited}, among the 22 submissions with citation counts exceeding 150, 14 appeared in the high pre-rebuttal score group, compared to 17 in the high-ranked group. In contrast, 6 of these well-cited submissions fell into the low pre-rebuttal score group, compared to 5 in the low-ranked group. Turning to final decisions, 15 of these submissions were accepted by ICML 2023, while 7 were rejected. Notably, 4 of the rejected submissions were nonetheless ranked as the top paper by at least one of their authors.


\section{Control for the confounding factors}
\label{sec:confounding}

Several potential confounders have already been addressed in our analyses. 
Figures~\ref{fig:norm_citation} and \ref{fig:CCDF_versus_citation} stratified submissions by final decisions, showing that within both accepted and rejected papers, higher-ranked submissions accrued substantially more citations. 
Figure~\ref{fig:rank_vs_score} further stratified by review scores, indicating that higher-ranked papers consistently received more citations even when controlling for similar scores. 
Our design also implicitly accounted for paper topics: for each author, the highest- and lowest-ranked papers were compared, thereby matching on author background and research area and (approximately) controlling for topic-related confounding.

To further validate previous findings, we examine additional potential confounders in this section. Publication time is one potential factor, as earlier arXiv postings tend to accumulate more citations. The upper panel of Figure~\ref{fig:confounder_check} shows the posting-time distributions for high- and low-ranked papers, with three key dates highlighted by red dashed lines. The distributions are similar, with mean posting dates of March 4, 2023 for high-ranked and March 8, 2023 for low-ranked papers, a difference of only four days.

Another factor is the authors' preferences. First, authors may self-cite high-ranked papers more frequently. To test this, we calculate citation counts excluding self-citations, which are defined as citations from papers with no overlapping authors. Among 649 authors with multiple ranked submissions and valid non-self-citation counts, the middle and lower panels of Figure~\ref{fig:confounder_check} present cumulative averages from July 23, 2023 to November 22, 2024, conditional on final decisions and review scores. The same conclusion holds: high-ranked papers continue to receive more citations even when self-citations are excluded.

Second, authors may devote greater effort to publicizing high-ranked papers at the conference. To assess this, we examine the two-month window beginning July 23, 2023, the opening of ICML 2023, shown in the lower panel of Figure~\ref{fig:norm_citation}. During this period, high-ranked submissions received an average of 1.62 citations, more than twice that of low-ranked submissions. While authors may promote their high-ranked papers during the conference, it typically takes months for new work to appear that cites them. Thus, the average citation counts observed between July 23 and September 22, 2023 are unlikely to be attributed to conference publicity. This provides further evidence that the strong association between author rankings and citation counts cannot be explained by promotion during the conference.

\section{Discussion}
\label{sec:discuss}

Recognizing that peer review is strained by a limited reviewer pool in the search for high-impact research, our study reveals that a powerful and overlooked source of insight is the authors themselves. We demonstrated through a large-scale experiment at one of the largest AI conferences that authors' comparative assessments of their own papers are a remarkably strong predictor of future scientific impact. Submissions that authors ranked highest garnered more than double the citations of those they ranked lowest; this signal was especially pronounced in the tail, identifying a large majority of the most highly cited papers. Crucially, self-rankings are more predictive of future citation counts than reviewers' scores. These findings are statistically robust and persisted after controlling for confounding factors like preprint posting dates and self-citations.

Citation counts are usually viewed as a strong indicator of a paper’s impact in AI. At AI conferences, papers typically introduce fresh ideas—either novel insights into existing models or entirely new methods—and there isn’t a single ``ground truth'' that immediately proves which approach is best; instead, the community tests, debates, and builds on them. In that context, a paper’s citation count serves as a practical signal of scientific impact: when many researchers cite a paper, it means the idea has been noticed, scrutinized, and found useful enough to inform further work, making it a reasonable proxy for scientific influence. Beyond academia, AI papers can also shape products and markets. Popular, highly cited ideas tend to attract developer mindshare, funding, and tooling—so the “hotter” the idea, the more people try it and the greater its perceived business value. That’s why heavily cited architectures (e.g., the Transformer \citep{vaswani2017attention}) have become the backbone of many commercial systems.

Beyond citation counts, the robustness of our findings is supported by analyses across multiple impact metrics. The predictive power of self-rankings held true not only for citations tracked by Semantic Scholar but also for Google Scholar citations and for a community engagement metric, GitHub stars. While these results are compelling, we acknowledge the need for validation over a longer time horizon and through experiments at more conferences to better assess the long-term impact of research work. To this end, we have already conducted similar experiments at ICML 2024 and 2025, as well as NeurIPS 2025, which we expect will validate and extend these findings. Further challenges include ensuring truthful reporting of self-rankings and determining how self-rankings can be formally incorporated into the review process. Notably, the comparative design of self-rankings renders it impossible for authors to simply inflate the quality of all their submissions. Indeed, an emerging body of literature is developing mechanisms in the game-theoretic framework that leverage rankings for more accurate review scores \citep{su2021you,yan2023isotonic}.

These results carry significant implications for the future of scholarly evaluation, particularly in fast-moving fields like AI. Such fields demand review systems capable of separating methodological correctness from likely scientific consequences. Current peer review in AI research tends to reward correctness and quantifiable improvements over broader impact. The increasing use of language models in the review process may further bias evaluations toward surface-level characteristics rather than profound scientific contributions. In contrast, our study demonstrates that self-rankings appear better aligned with anticipated field-shaping value, perhaps because authors possess a deeper, more holistic understanding of their work's conceptual novelty and long-term potential. This information is particularly valuable in the selection of best paper awards at AI conferences, which are intended to promote high-impact research. Our work provides compelling evidence that authors' self-rankings are a low-cost, reliable signal that can be incorporated into future peer review systems. Indeed, based on the findings from the ICML 2023 experiment, ICML 2026 will, for the first time, incorporate author self-rankings into its formal decision-making process (see \url{https://icml.cc/Conferences/2026/CallForPapers}).

\subsection*{Methods}
Methods are available in the Section \ref{sec:method}.

\subsection*{Acknowledgments}
We would like to thank Francis Bach, Melisa Bok, Emma Brunskill, Barbara Engelhardt, Sherry Xue, and James Zou for helpful discussions. We are grateful to the ICML Board for providing the opportunity for this experiment. This research was supported in part by Analytics at Wharton, Wharton AI \& Analytics Initiative, the National Institute of Mental Health under Award Number R01MH136055, and National Institute on Aging under Award Numbers RF1AG082938 and R01AG085581. The content is solely the responsibility of the authors and does not necessarily represent the official views of the National Institutes of Health. 



\bibliographystyle{plainnat}
\bibliography{ref}

\clearpage

\appendix

\setcounter{figure}{0}
\setcounter{table}{0}

\renewcommand{\thefigure}{S.\arabic{figure}}
\renewcommand{\thetable}{S.\arabic{table}}

\section{Methods}
\label{sec:method}

\subsection{Survey Question}

\begin{figure}[!htp]
    \centering
    \includegraphics[width=0.8\textwidth]{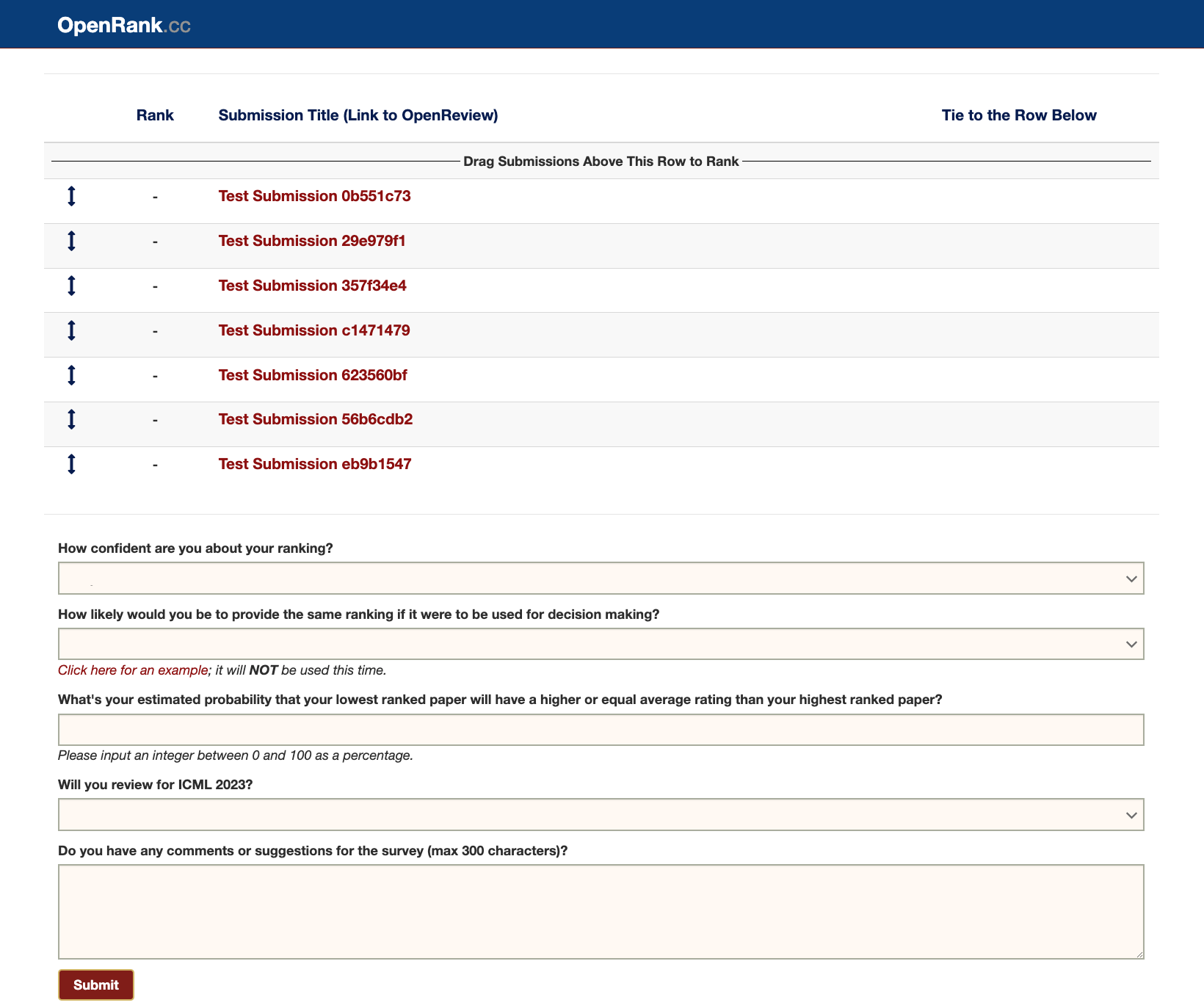}
    \caption{Screenshot of the first author survey. This survey was sent out on January 26, 2023, and closed on February 10, 2023. Among the 1,342 authors with multiple submissions who provided valid rankings, they took an average of around 4 days to submit the survey: 25\% of them submitted the survey within 6.21 hours and 90\% of them submitted the survey within 8.79 days.}
    \label{fig:survey_interface}
\end{figure}

Figure \ref{fig:survey_interface} presents full screenshots of the surveys. Authors with multiple submissions were asked to rank their papers based on their perceived quality by dragging them up or down. 
All authors were asked to respond to the questions shown in Figure \ref{fig:survey_interface}.
A comprehensive analysis of survey questions other than the self-ranking is presented in \citep{su2025icml}.

\subsection{Extract Citation Counts}

To obtain the citation counts for each submission from Semantic Scholar, we followed a systematic procedure. 
First, we input each submission title through Semantic Scholar API to identify the best-matching record. 
Second, we extract the submission title, author list from the best-matching record. 
Third, we employed the Levenshtein distance metric to ensure the submission found on Semantic Scholar matches the corresponding ICML 2023 submission. Intuitively, the Levenshtein distance measures the difference between two sequences by calculating the minimum number of single-character edits (insertions, deletions, or substitutions) required to transform one string into another. 
If both the title and author list of the best-matching record have a Levenshtein distance of 20 or less from those of the ICML 2023 submission, we consider them to refer to the same paper. 
For instance, we allow minor title changes, such as replacing one or two words, or modifications to the author list, such as the addition of middle names not recorded in OpenReview or the inclusion of a newly added co-author in subsequent versions of the paper.
We also allow the paper title to include special string like ``Schrödinger'', which may cause minor mismatch in downloading the title from Semantic Scholar.




\section{Additional analyses}

\subsection{Additional details of Section \ref{par:rank_vs_avg_ciation}-\ref{sec:rank_vs_score}}

\begin{figure}[!htp]
    \centering
    \begin{subfigure}[b]{0.43\textwidth}
        \includegraphics[height=\textwidth]{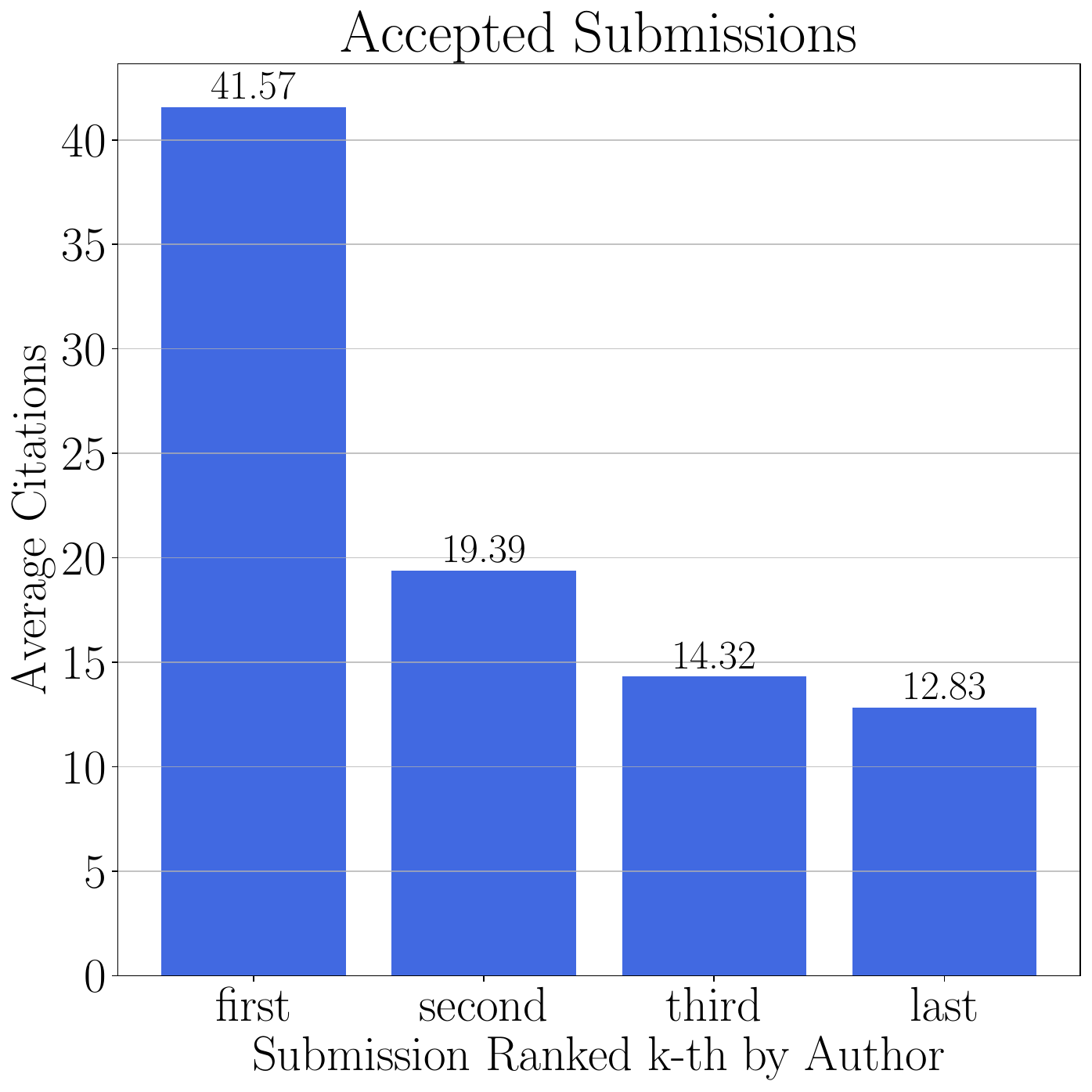}
    \end{subfigure}
    \begin{subfigure}[b]{0.43\textwidth}
        \includegraphics[height=\textwidth]{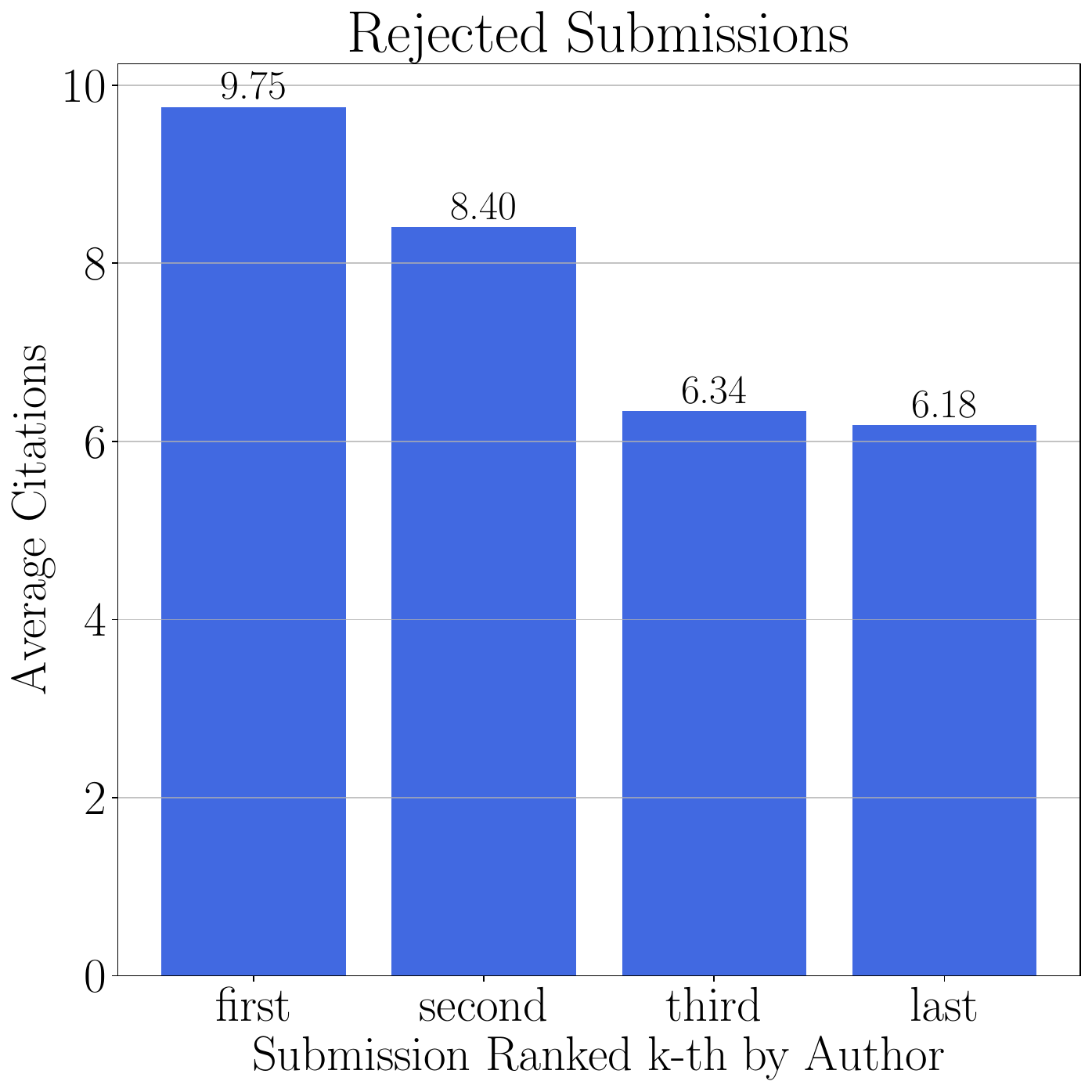}
    \end{subfigure}
    \caption{Citation counts from  from July 23, 2023, to November 22, 2024, for ranked first, second, third and last papers, grouped by final acceptance or rejection decisions at ICML 2023. High ranked papers receive significantly more citations.}
    \label{fig:top_k_citation}
\end{figure}

Figure~\ref{fig:top_k_citation} extends our analysis beyond the comparison of only the top- and bottom-ranked submissions. We consider 76 authors with more than three accepted submissions and 86 authors with more than three rejected submissions, all of whom provided valid citation data. For each author, we classify their submissions into four groups based on their rankings: first, second, third, and last. As shown in Figure~\ref{fig:top_k_citation}, low-ranked submissions tend to accumulate fewer citations, suggesting that self-rankings, even beyond the top and bottom positions, carry meaningful information about paper impact.

Table~\ref{tab:citation_versus_rank} reports the citation counts for high- versus low-ranked submissions, conditional on their final decisions. These results complement Figure~\ref{fig:CCDF_versus_citation} by providing exact statistics.

\begin{table}[!htp]
\centering
\renewcommand{\arraystretch}{1.2}
\resizebox{\textwidth}{!}{
\begin{tabular}{l||C{3cm}|C{3cm}|C{3cm}|C{3cm}}

\hline \hline
\multirow{2}{*}{} & \multicolumn{2}{c|}{Citation Count for Accepted Papers} & \multicolumn{2}{c}{Citation Count for Rejected Papers} \\

\cline{2-5}
& \begin{tabular}{@{}c@{}} high-ranked \end{tabular} & \begin{tabular}{@{}c@{}} low-ranked \end{tabular} & \begin{tabular}{@{}c@{}} high-ranked \end{tabular} & \begin{tabular}{@{}c@{}} low-ranked \end{tabular}
\\

\hline
\begin{tabular}{@{}c@{}} Mean \end{tabular} & $27.36$ & $13.82$ & $12.41$ & $6.36$ \\ 

\hline
\begin{tabular}{@{}c@{}} $p$-value \end{tabular} & \multicolumn{2}{c|}{$9.34 \times 10^{-3}$} & \multicolumn{2}{c}{$7.68 \times 10^{-4}$} \\

\hline
\begin{tabular}{@{}c@{}} Max \end{tabular} & 979.0 & 331.0 & 253.0 & 205.0 \\ 

\hline
\begin{tabular}{@{}c@{}} $75$th percentile \end{tabular} & 21.0 & 15.0 & 8.0 & 5.0 \\ 

\hline
\begin{tabular}{@{}c@{}} $50$th percentile \end{tabular} & 8.0 &  6.0 & 3.0 & 2.0 \\ 

\hline
\begin{tabular}{@{}c@{}} $25$th percentile \end{tabular} & 3.0 & 3.0 & 1.0 & 0.0 \\ 

\hline\hline
\begin{tabular}{@{}c@{}} Number of authors \end{tabular} & \multicolumn{2}{c|}{$280$} & \multicolumn{2}{c}{$343$} \\ 

\hline
\end{tabular}}
\caption{Citation counts from July 23, 2023, to November 22, 2024, for high- versus low-ranked papers, grouped by final acceptance or rejection decisions at ICML 2023. A paired $t$-test shows that, at the $p = 0.05$ significance level, high-ranked papers receive significantly more citations than low-ranked papers.}
\label{tab:citation_versus_rank}
\end{table}

\begin{figure}[!htp]
    \centering
    \begin{subfigure}[b]{0.43\textwidth}
        \includegraphics[width=\textwidth]{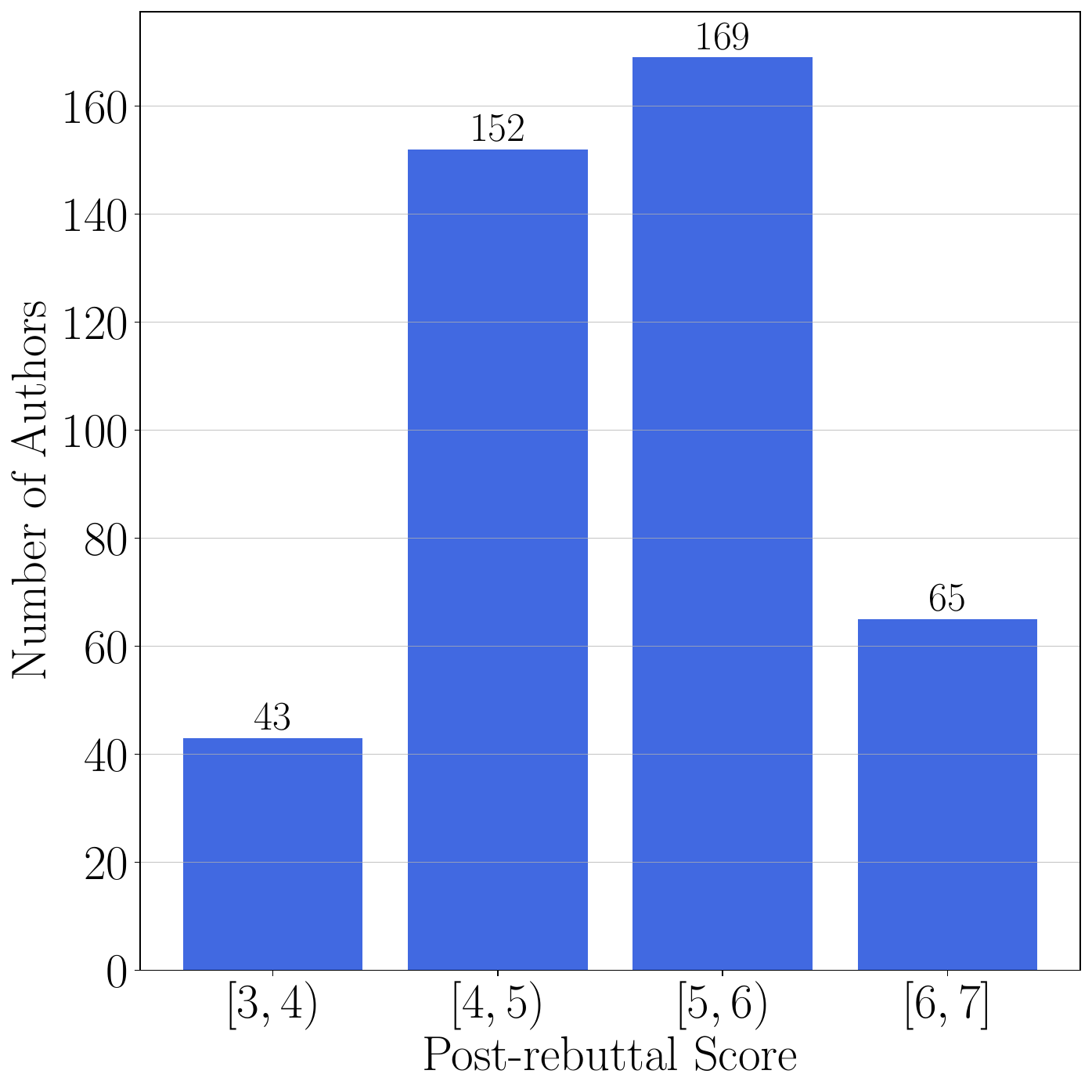}
    \end{subfigure}
    \begin{subfigure}[b]{0.43\textwidth}
        \includegraphics[width=\textwidth]{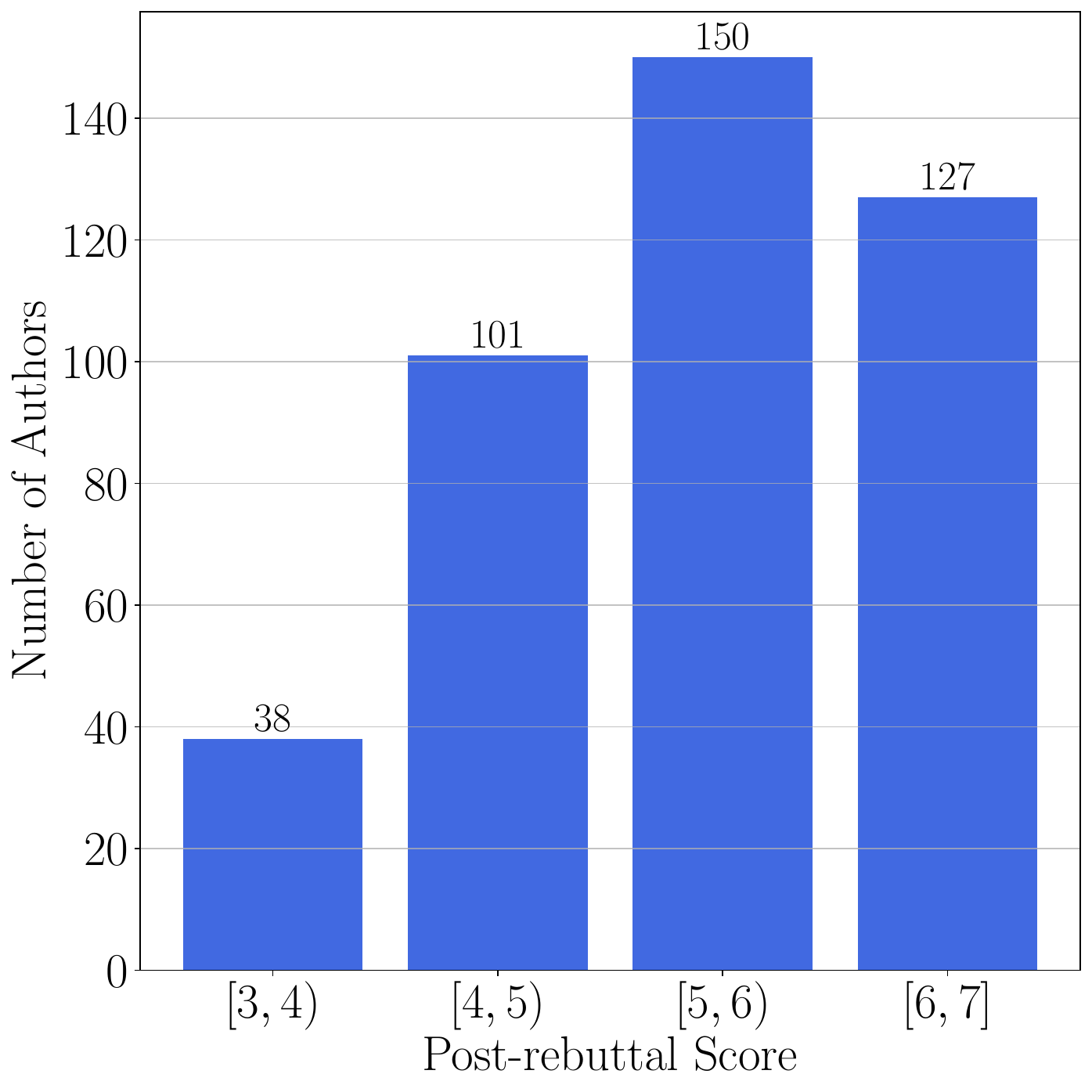}
    \end{subfigure}
    \caption{Number of authors with multiple ranked submissions whose review scores fall into each score interval.}
    \label{fig:summery_cond_score}
\end{figure}

Figure~\ref{fig:summery_cond_score} reports the number of authors with multiple ranked submissions, grouped by the review scores of their papers. This serves as a summary of the sample sizes for each group shown in Figure~\ref{fig:rank_vs_score}.

\subsection{Additional details of Section \ref{sec:confounding}}

\begin{figure}[!htp]
    \centering
    \makebox[\textwidth][l]{
        \hspace{0.01\textwidth}
        \begin{subfigure}[b]{0.46\textwidth}
            \includegraphics[height=0.8\textwidth]{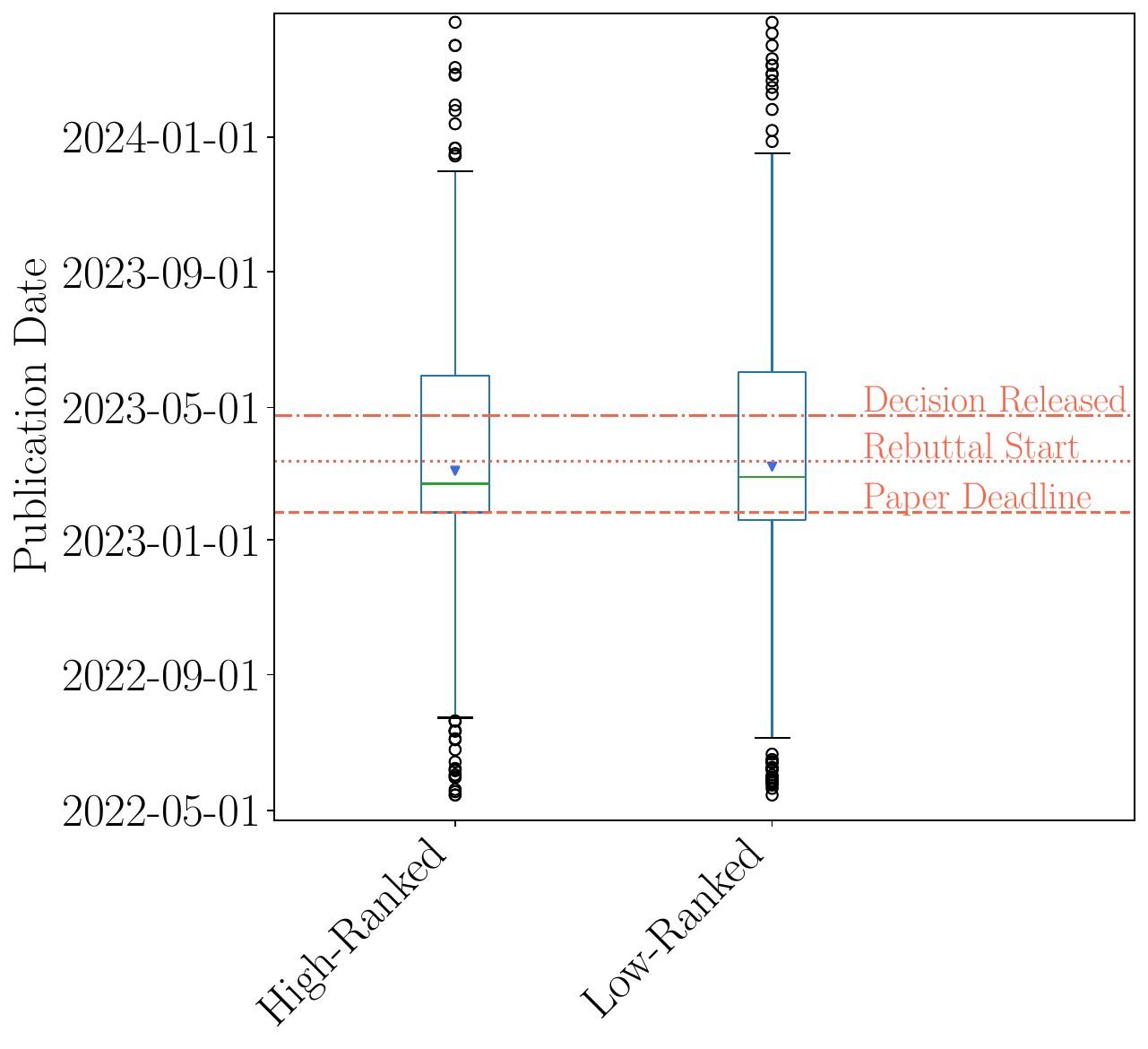}
        \end{subfigure}
        \hspace{-0.05\textwidth}
        \begin{subfigure}[b]{0.46\textwidth}
            \includegraphics[height=0.8\textwidth]{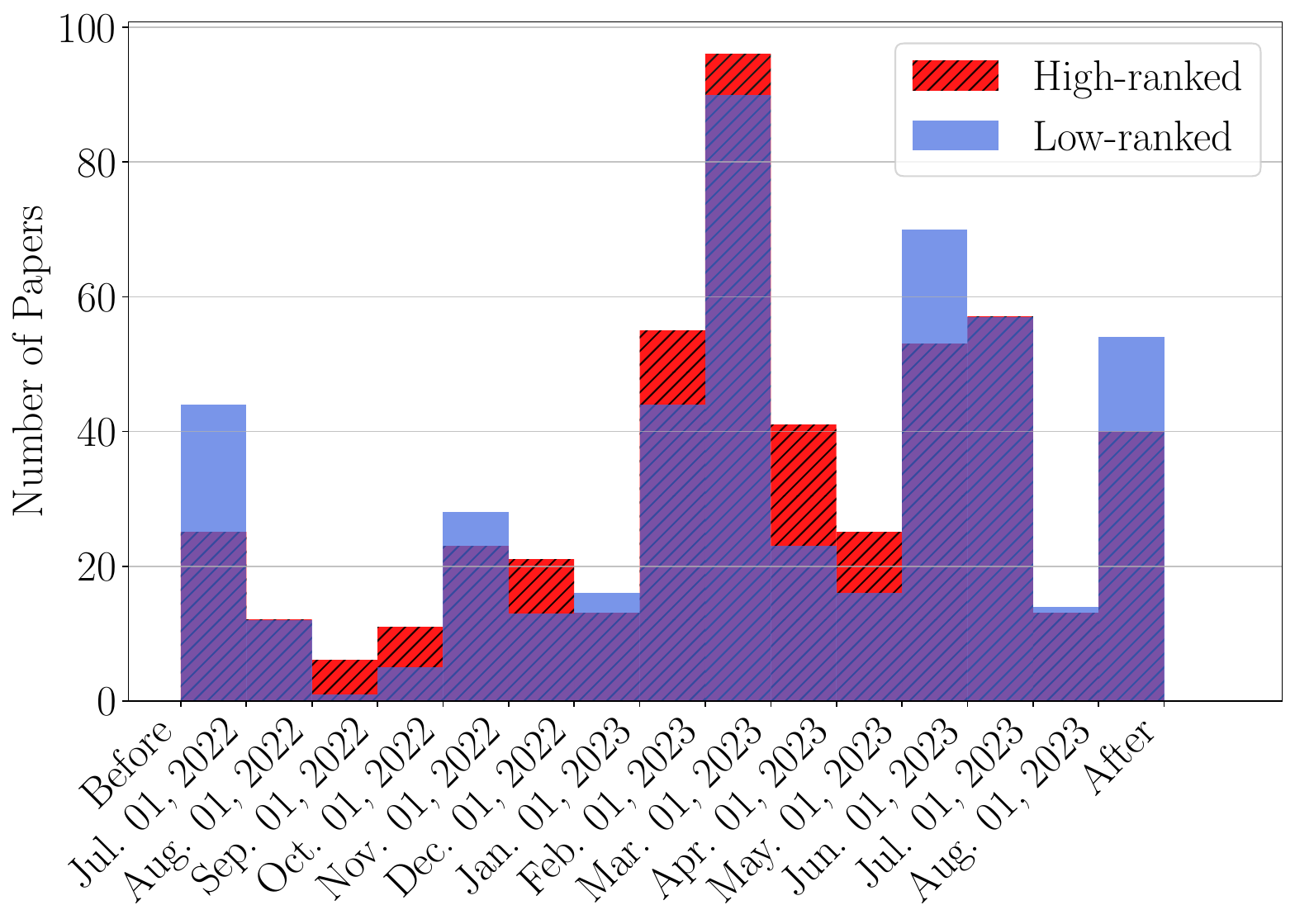}
        \end{subfigure}
    }
    
    \vspace{0.1em}
    
    \begin{subfigure}[b]{0.465\textwidth}
        \includegraphics[height=0.83\textwidth]{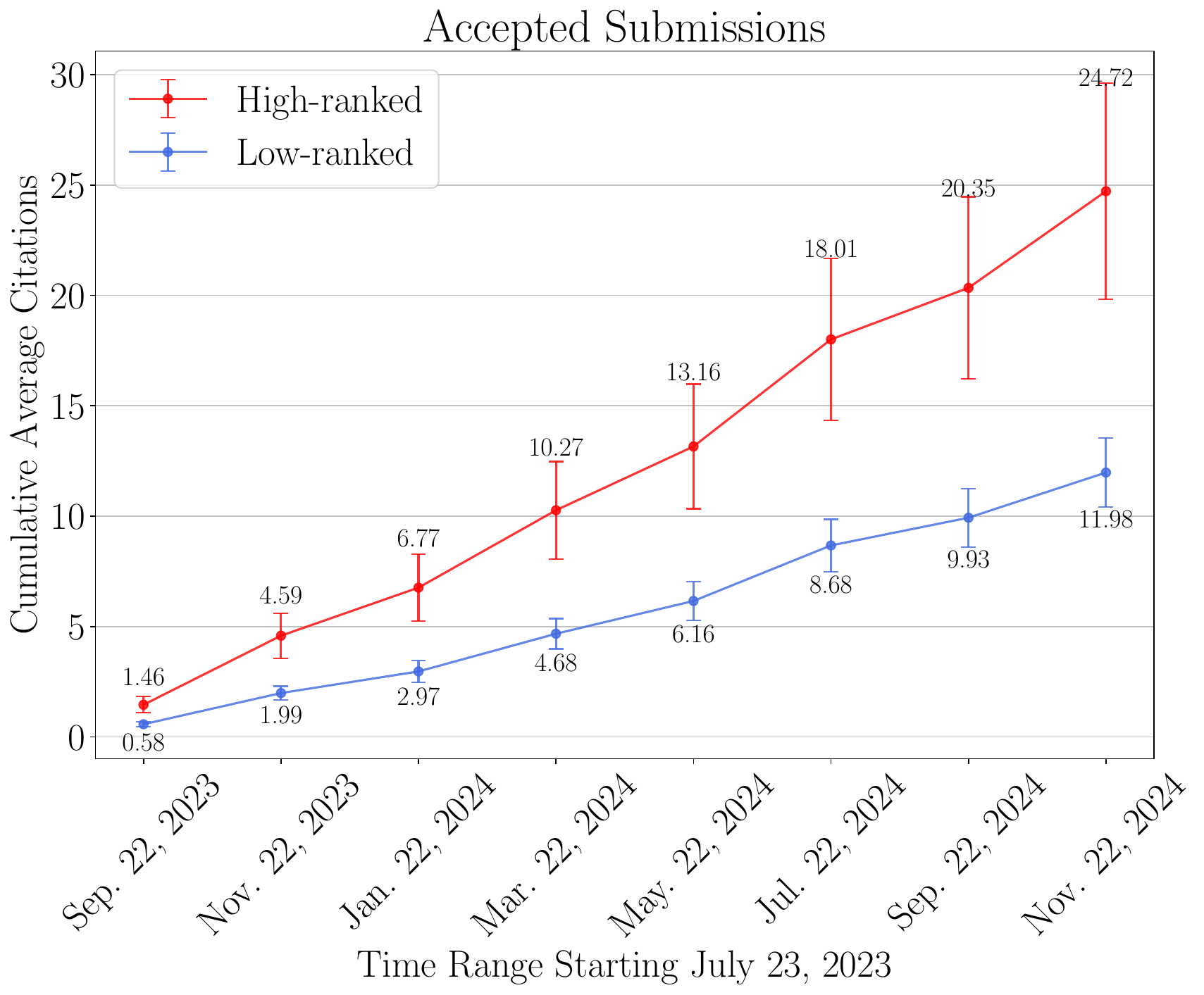}
    \end{subfigure}
    \begin{subfigure}[b]{0.465\textwidth}
        \includegraphics[height=0.83\textwidth]{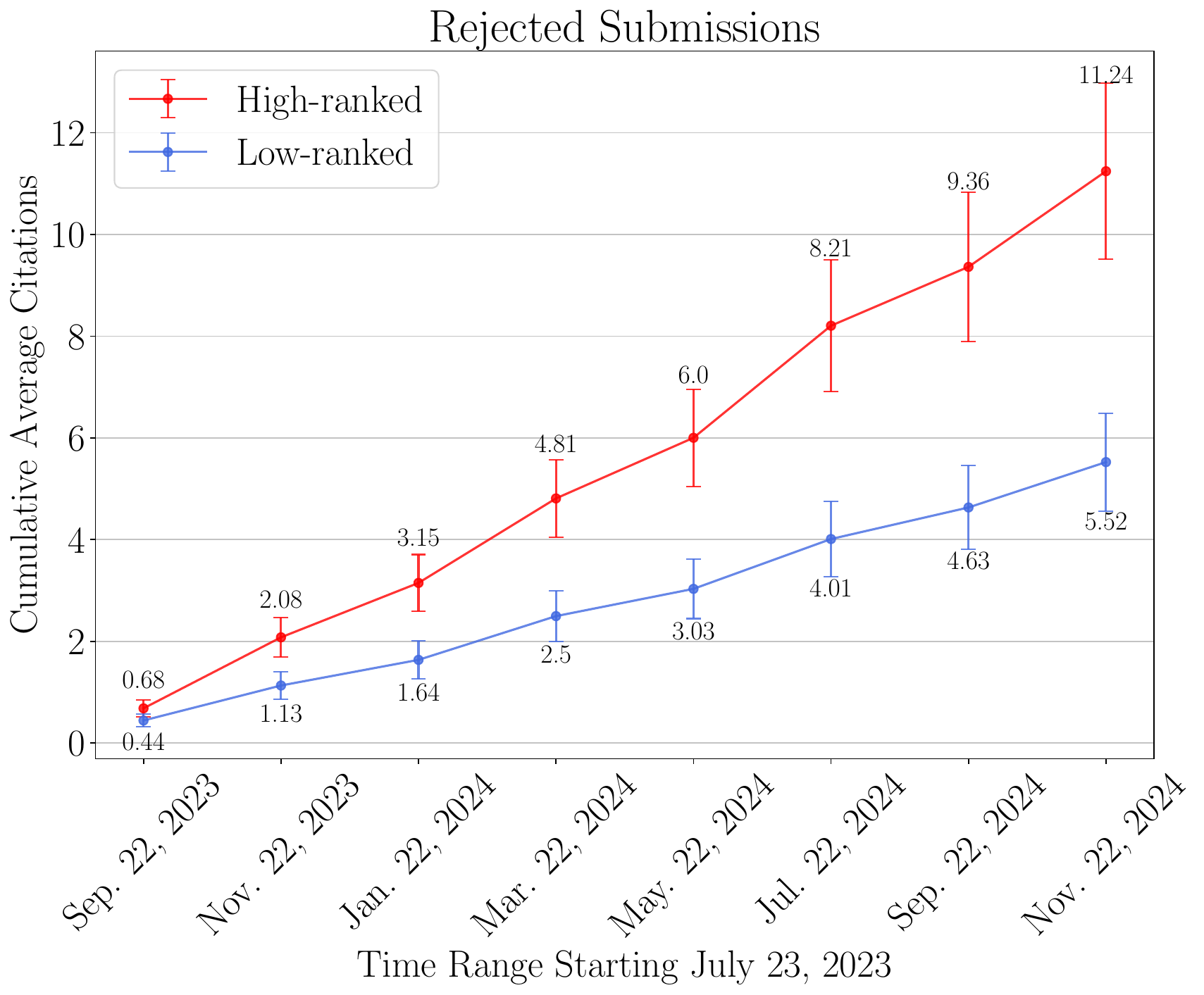}
    \end{subfigure}

    \vspace{0.1em}
    
    \begin{subfigure}[b]{0.465\textwidth}
        \includegraphics[width=\textwidth]{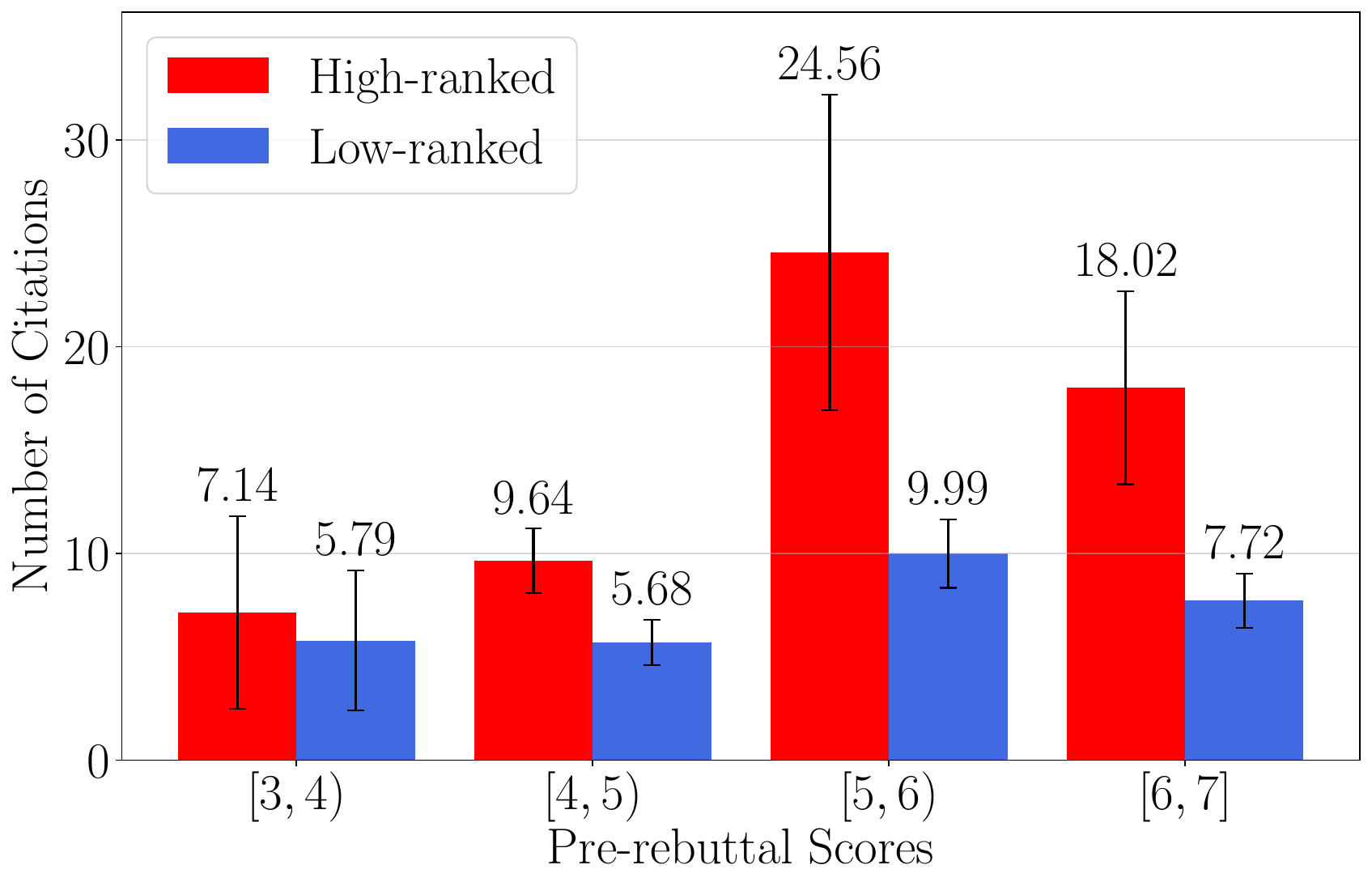}
    \end{subfigure}
    \begin{subfigure}[b]{0.465\textwidth}
        \includegraphics[width=\textwidth]{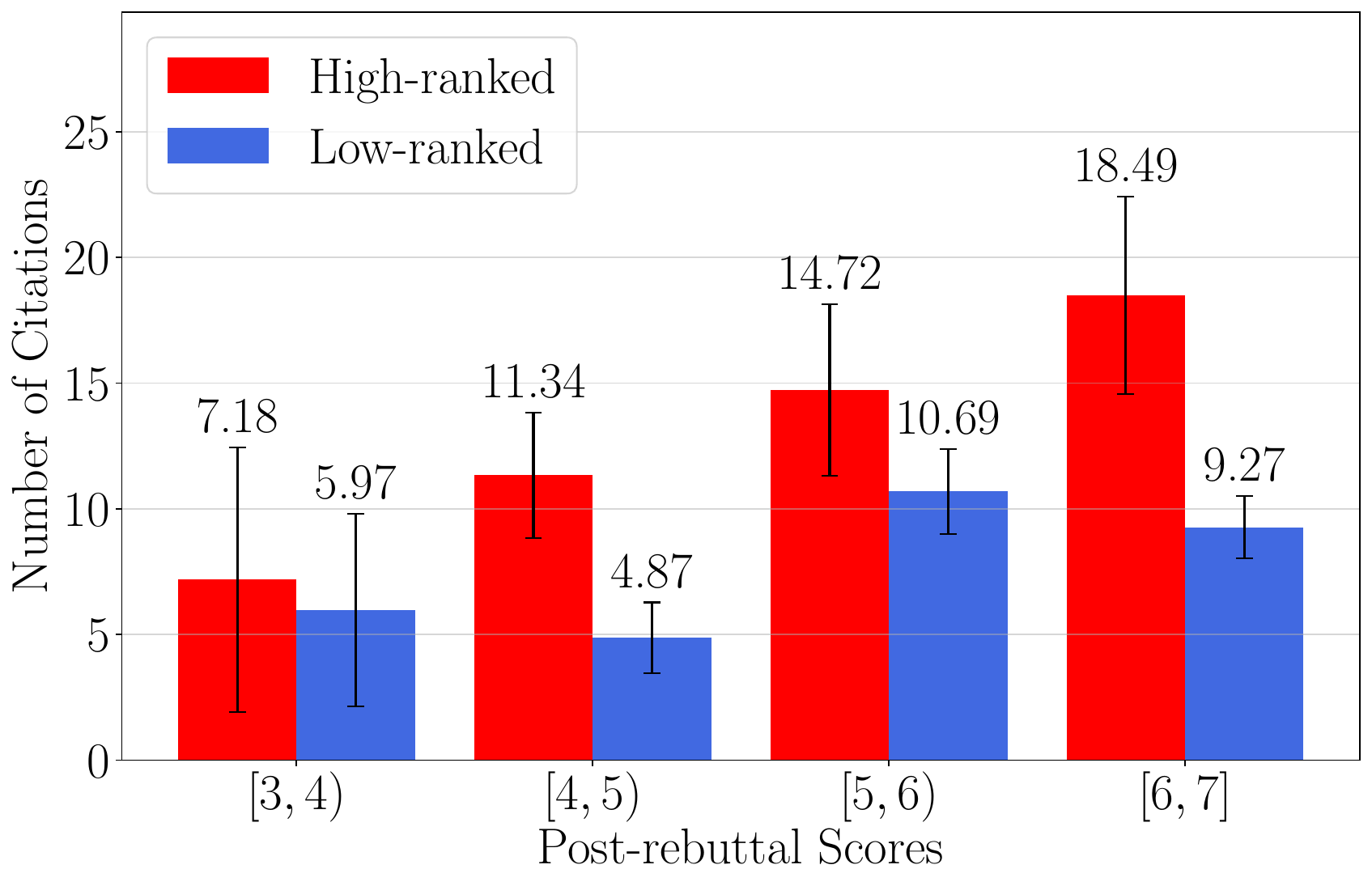}
    \end{subfigure}
    \caption{{High-ranked papers are posted to arXiv at similar times as low-ranked ones. Moreover, previous conclusions remain after removing self-citations.} 
    {Upper panel}: Distribution of arXiv posting dates for high- and low-ranked papers. The mean posting dates differ by only 4 days, and a Welch’s $t$-test yields a $p$-value of $0.72$, indicating no significant difference. Thus, arXiv timing is unlikely to confound our findings.  Red dashed lines indicate three important date at ICML 2023: ``Paper Decision Notification,'' ``Review Release to Authors,'' and ``Full Paper Submission Deadline.'' 
    {Middle and Lower panels}: Replication of Figures \ref{fig:norm_citation} and \ref{fig:rank_vs_score}, excluding self-citations between July 23, 2023, and November 22, 2024.}
    \label{fig:confounder_check}
\end{figure}

Figure~\ref{fig:confounder_check} presents the arXiv posting dates for high- and low-ranked papers, along with replications of Figures~\ref{fig:norm_citation} and \ref{fig:rank_vs_score}, excluding self-citations between July 23, 2023, and November 22, 2024.
We observe that the distribution of arXiv posting times is similar between the high- and low-ranked groups, suggesting that the timing of arXiv release is reasonably balanced. Moreover, our main results continue to hold after excluding self-citations.

We also note that, in the upper right panel of Figure~\ref{fig:confounder_check}, to quantify the posting time, we converted each arXiv posting datetime into the number of seconds since January 1, 1970 at 00:00:00 UTC, commonly known as the UNIX timestamp.
A Welch's $t$-test comparing the UNIX timestamp in two groups yields a $p$-value of $0.72$, suggesting that arXiv posting time does not significantly differ between groups and is unlikely to confound our analysis.
Moreover, higher proportion of the low-ranked papers posted either very early or very late compared to the high-ranked papers. 
This could be attributed to differences in submission quality. 
Low-ranked papers are more likely to have been rejected from prior venues and later recycled for submission to ICML 2023 (see Table 1 in \cite{su2025icml}). Similarly, papers posted very late on arXiv often correspond to submissions that were rejected from ICML 2023 and later resubmitted to subsequent AI conferences.
Finally, we make a observation related to the adversarial aspects of the double-blind review process. Notably, over 25\% of submissions were posted on arXiv before the paper submission deadline, and over 50\% were posted prior to the review release to authors. Such early postings may introduce bias into the review process and pose challenges to maintaining authors' anonymity.

\subsection{Additional details of Section \ref{sec:discuss}}
\label{sec:other_metric}

To further examine how self-rankings align with paper quality and impact, we consider GitHub Stars as an alternative metric. We compare high- and low-ranked papers in terms of the GitHub Stars they received.

\paragraph{GitHub stars.}

Among the 2,530 submissions with valid rankings, 678 submissions were matched to a GitHub repository with valid GitHub Star counts. This set includes 232 authors with multiple ranked submissions and valid GitHub data.
Using the same grouping approach as before, we find that high-ranked and accepted papers received an average of 755.15 GitHub Stars, more than twice that of low-ranked and accepted papers, which averaged 347.48 Stars.
A similar trend is observed among rejected submissions. 
High-ranked and rejected papers received 89.77 GitHub Stars on average, while low-ranked and rejected papers received 33.27 Stars.

\paragraph{Google Scholar citations.}

\begin{figure}[!htp]
    \centering
    \begin{subfigure}[b]{0.43\textwidth}
        \includegraphics[width=\textwidth]{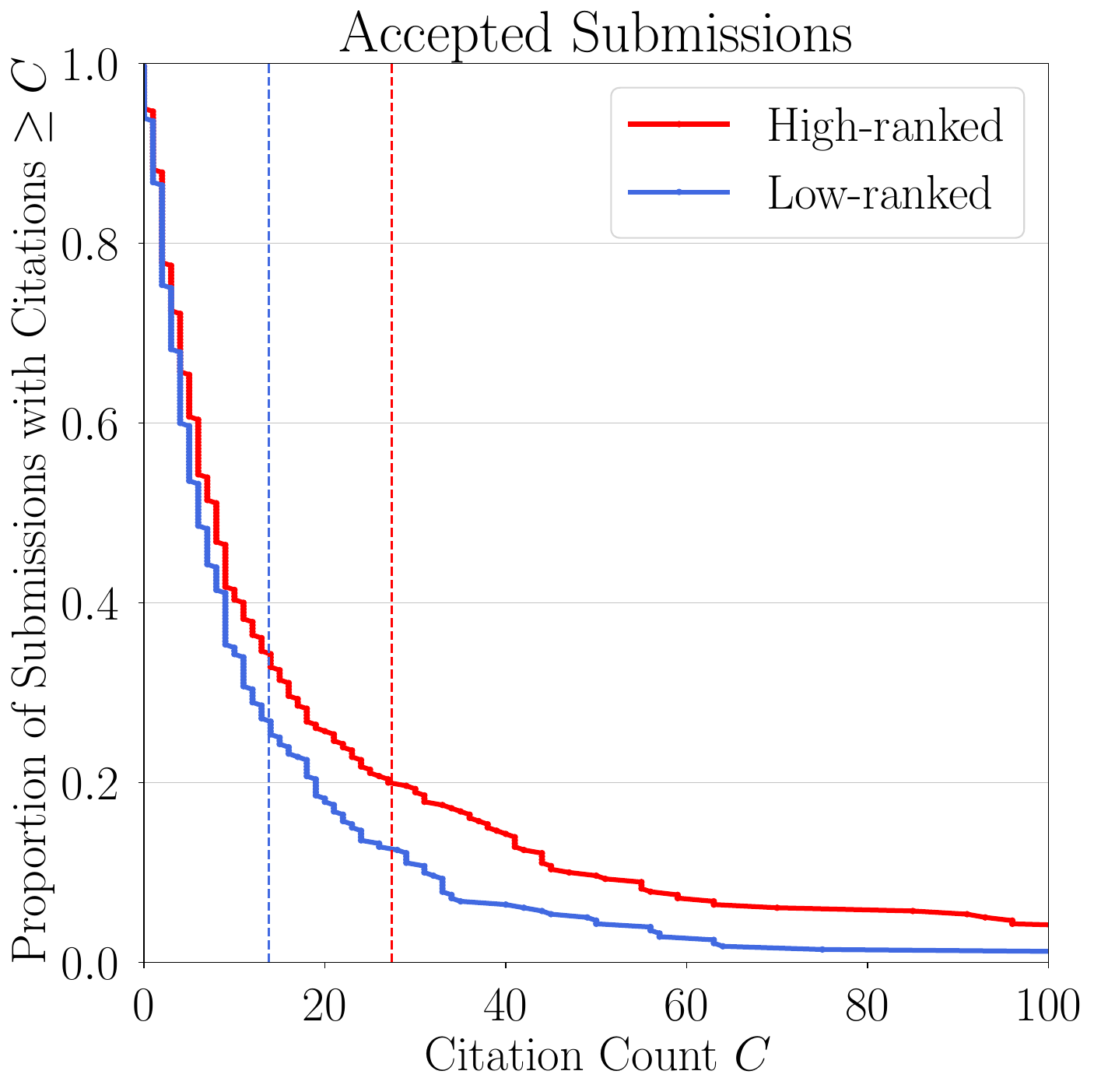}
    \end{subfigure}
    \begin{subfigure}[b]{0.43\textwidth}
        \includegraphics[width=\textwidth]{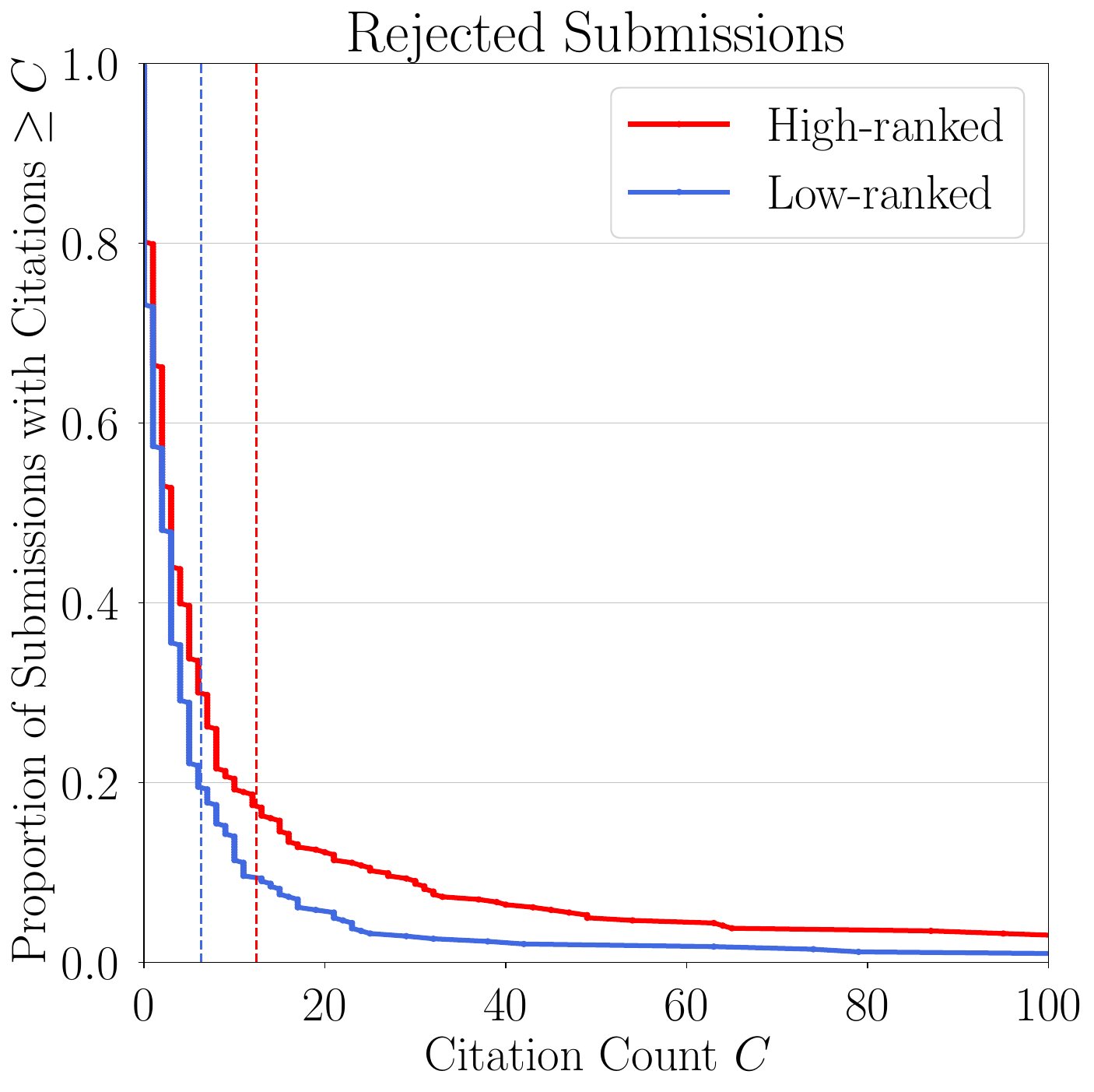}
    \end{subfigure}
    \caption{Empirical complementary cumulative distribution of citation counts, showing the proportion of papers with more than $x$ citations. The left panel shows results for accepted submissions, while the right panel presents results for rejected submissions. Across all thresholds, high-ranked papers consistently exhibit higher citation counts than low-ranked papers.}
    \label{fig:CCDF_versus_google_citation}
\end{figure}

\begin{figure}[!htp]
    \centering
    \begin{subfigure}[b]{0.49\textwidth}
        \includegraphics[width=\textwidth]{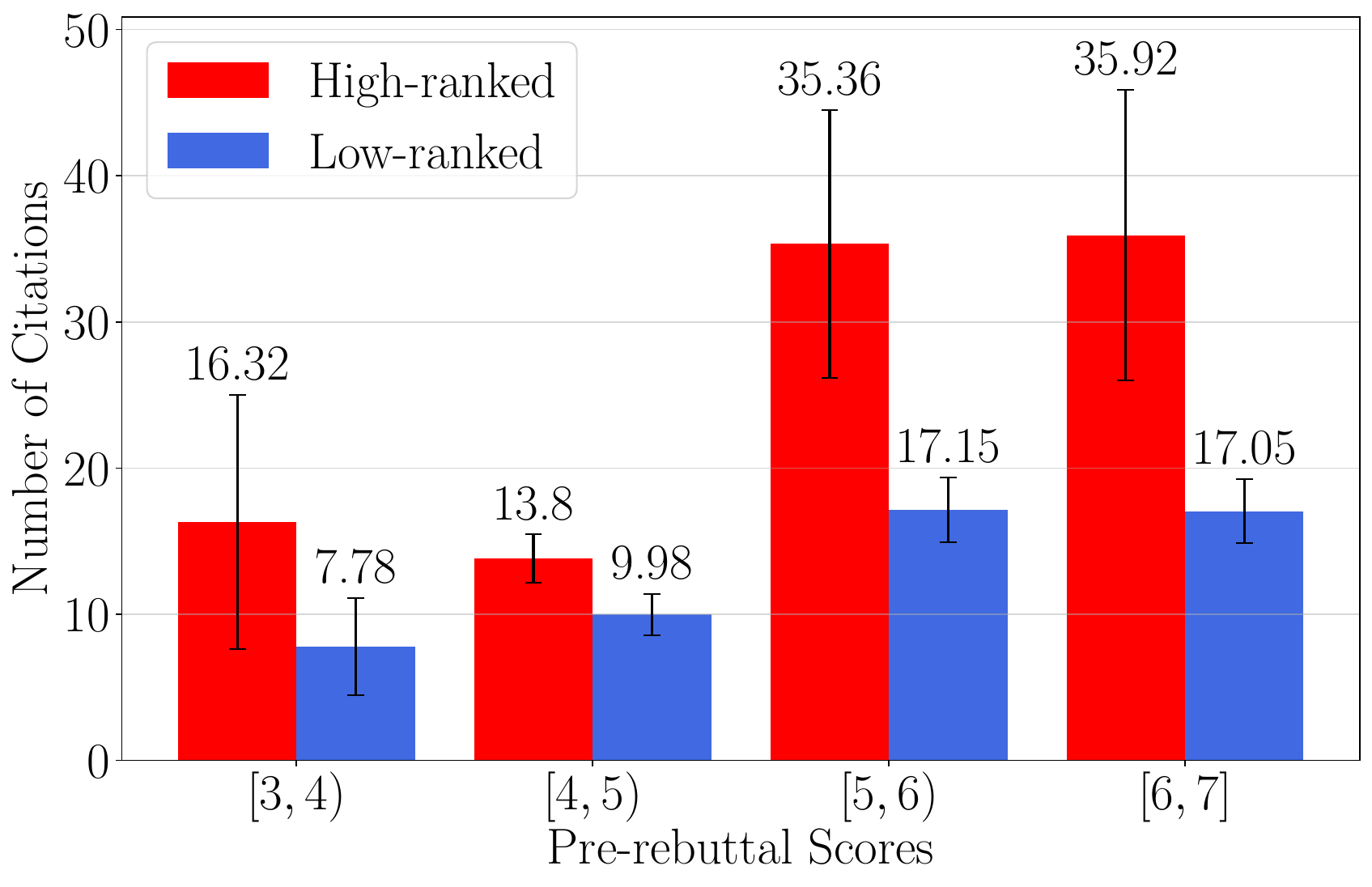}
    \end{subfigure}
    \begin{subfigure}[b]{0.49\textwidth}
        \includegraphics[width=\textwidth]{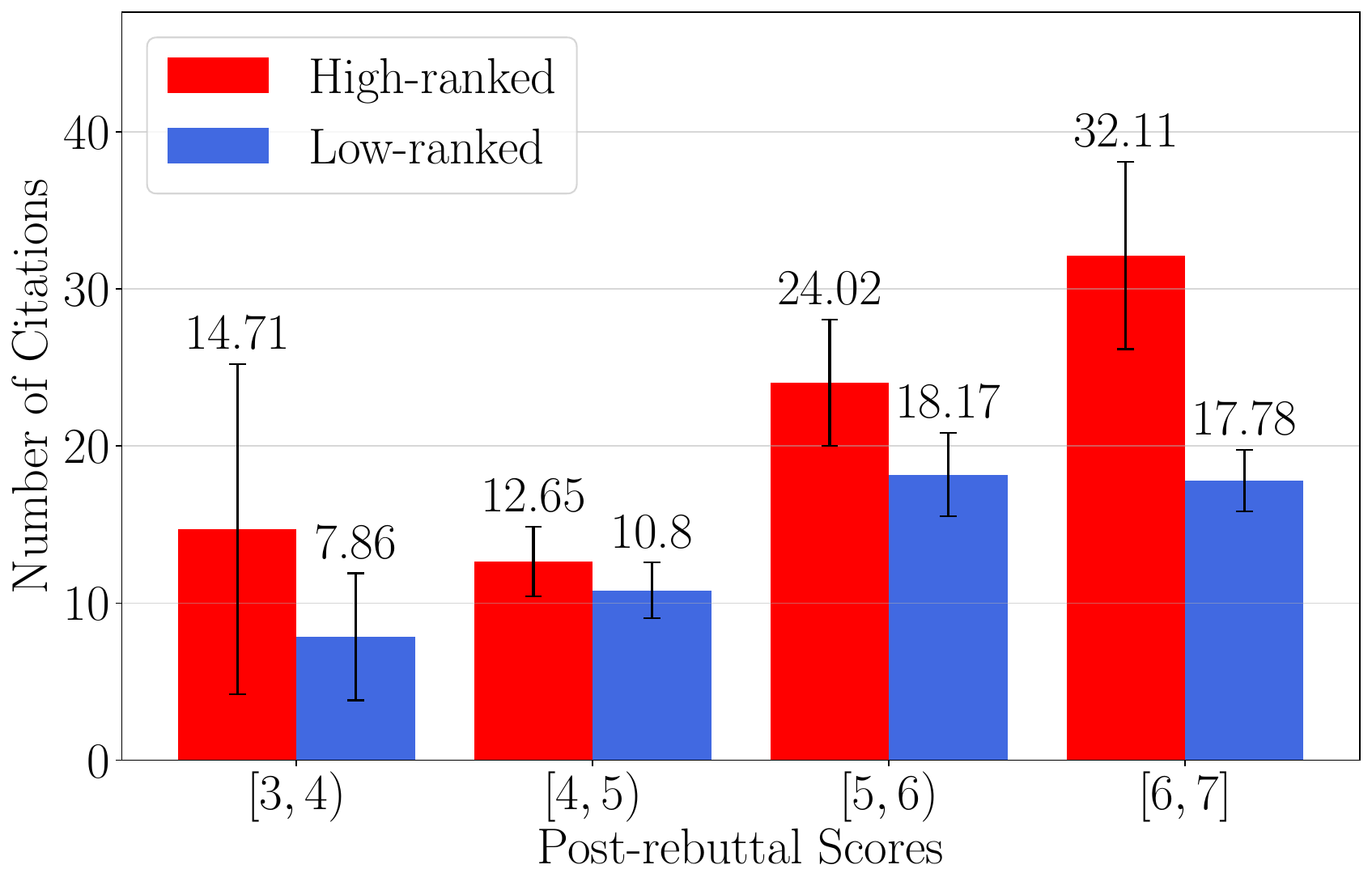}
    \end{subfigure}
    \caption{Average number of Google Scholar citations from July 23, 2023 to November 22, 2024 for high- vs. low-ranked papers, grouped by pre-rebuttal and post-rebuttal review scores at ICML 2023. Among submissions with similar review scores, high-ranked papers still receive significantly more citations than low-ranked papers.}
    \label{fig:rank_vs_score_google}
\end{figure}

We further validate the findings in Section~\ref{par:rank_idfy_good}-\ref{sec:rank_vs_score} by repeating the analysis using citation counts from Google Scholar. Since Google Scholar does not support filtering by date range and does not exclude self-citations, we restrict the analysis to those presented in Table~\ref{tab:rank_vs_score}, and Figures~\ref{fig:CCDF_versus_citation} and~\ref{fig:rank_vs_score}.
The corresponding results using Google Scholar are reported in Table~\ref{tab:rank_vs_score_google}, Figures~\ref{fig:CCDF_versus_google_citation} and~\ref{fig:rank_vs_score_google}. We find that the patterns are highly similar to those observed using citation counts from Semantic Scholar.
One notable difference is that the citation counts in Table~\ref{tab:rank_vs_score_google} are higher than those in Table~\ref{tab:rank_vs_score}. This is expected, as the Google Scholar counts are aggregated from the paper’s initial arXiv posting up to November 22, 2024, while the Semantic Scholar data in Table~\ref{tab:citation_versus_rank} only includes citations from July 23, 2023 to November 22, 2024.


\begin{table}[!htp]
\centering
\renewcommand{\arraystretch}{1.2}
\resizebox{\textwidth}{!}{
\begin{tabular}{>{\centering\arraybackslash}m{1.7cm}|p{1.3cm}||c|c|c|c|c|c}
\hline \hline
 & & \multicolumn{3}{c}{Citation Count for Accepted Papers} & \multicolumn{3}{|c}{Citation Count for Rejected Papers} \\
\cline{3-5} \cline{6-8}
 & & \begin{tabular}{@{}c@{}} Rank \end{tabular} & \begin{tabular}{@{}c@{}} Post-Rebuttal \\ Score \end{tabular} & 
\begin{tabular}{@{}c@{}} Pre-Rebuttal \\ Score \end{tabular} & \begin{tabular}{@{}c@{}} Rank \end{tabular} & \begin{tabular}{@{}c@{}} Post-Rebuttal \\ Score \end{tabular} & 
\begin{tabular}{@{}c@{}} Pre-Rebuttal \\ Score \end{tabular} \\
\hline
\multirow{2}{*}{\centering Mean} & High & 41.57 & 39.20 & 36.13 & 16.40 & 14.85 & 15.35 \\
\cline{2-8}
 & Low  & 21.25 & 26.46 & 27.97 & 10.24 & 13.54 & 11.55 \\
\hline
\multicolumn{2}{c||}{$p$-value}  & $6.3 \times 10^{-3}$ & $0.10$ & $0.29$ & $1.0 \times 10^{-2}$ & $0.67$ & $0.18$ \\
\hline
\multirow{2}{*}{\centering Max} & High & 1462.0 & 880.0 & 880.0 & 539.0 & 516.0 & 539.0 \\
\cline{2-8}
 & Low  & 485.0 & 1462.0 & 1462.0 & 516.0 & 539.0 & 516.0 \\
\hline
\multirow{2}{*}{\centering \begin{tabular}{@{}c@{}} 75th \\ percentile \end{tabular}} & High & 33.5 & 33.5 & 31.5 & 12.0 & 10.0 & 11.0 \\
\cline{2-8}
 & Low  & 24.0 & 25.5 & 26.0 & 9.0 & 10.0 & 9.0 \\
\hline
\end{tabular}}
\caption{Average number of Google Scholar citations (from July 23, 2023, to November 22, 2024) for (1) high-ranked versus low-ranked papers, (2) high-scored versus low-scored paper in post-rebuttal phase, and (3) high-scored versus low-scored paper in pre-rebuttal phase, grouped by their final decisions in ICML 2023. At a significance level of $p = 0.05$, only the high-ranked papers show a significantly higher number of citations compared to the low-ranked papers, based on a paired $t$-test.}
\label{tab:rank_vs_score_google}
\end{table}

\end{document}